\documentclass[sn-matphys,iicol,nofootinbib]{sn-jnl}
\usepackage[sort&compress,numbers]{natbib}
\usepackage{multirow}
\usepackage{graphicx}
\usepackage{amssymb}
\usepackage{amsmath}
\usepackage[full]{textcomp}%
\usepackage{booktabs}%
\usepackage[svgnames]{xcolor}
\usepackage{mathtools,slashed}
\usepackage[retainorgcmds]{IEEEtrantools}
\usepackage{physics}
\usepackage{epstopdf}
\usepackage[utf8]{inputenc}
\usepackage{subfig}
\usepackage{doi}
\usepackage{hyperref}
\usepackage{soul}
\usepackage{newtxtext}
\usepackage{newtxmath}

\usepackage{epsfig}

\newcommand{\be}{\begin{eqnarray}}
\newcommand{\ee}{\end{eqnarray}}
\begin{document}

\title{Collision energy dependence of source sizes for primary and secondary pions at energies available at the JINR Nuclotron-based Ion Collider
fAcility from L\'evy fits}

\author[1]{Alejandro Ayala}
\author[1]{Santiago Bernal-Langarica}
\author[2]{Isabel Dominguez}
\author[3]{Ivonne Maldonado}
\author[4]{Maria Elena Tejeda-Yeomans}
  \affil[1]{
  \orgdiv{Instituto de Ciencias
  Nucleares}, \orgname{Universidad Nacional Aut\'onoma de M\'exico}, \orgaddress{\postcode{Apartado
  Postal 70-543}, \city{CdMx 04510},
  \country{Mexico}}}
  \affil[2]{\orgdiv{Facultad de Ciencias Físico-Matemáticas}, \orgname{Universidad Autónoma de Sinaloa}, \orgaddress{\street{Avenida de las Américas y Boulevard Universitarios}, \city{Ciudad Universitaria, Culiacán}, \postcode{80000}, \country{Mexico}}}
  \affil[3]{\orgname{Joint Institute for Nuclear Research}, \orgaddress{\city{Dubna}, \postcode{141980}, \country{Russia}}}
  \affil[4]{\orgdiv{Facultad de Ciencias-CUICBAS}, \orgname{Universidad de Colima}, \orgaddress{\street{Bernal Díaz del Castillo No. 340, Colonia Villas San Sebastián}, \city{Colima}, \postcode{28045}, \country{Mexico}}}

\maketitle

\begin{abstract}

We study the evolution of the two-pion correlation function parameters with collision energy in the context of relativistic heavy-ion collisions within the NICA energy range. To this end, we perform UrQMD simulations in the cascade mode to produce samples of pions from $5\times 10^6$ Bi+Bi collisions for each of the studied energies. The effects of the quantum-statistical correlations are introduced using the correlation afterburner code CRAB. We fit the correlation function using Gaussian, exponential and symmetric L\'evy shapes and show that for all collision energies the latter provides the best fit. We separate the sample into pions coming from primary processes and pions originating from the decay of long-lived resonances, and show that the source size for the latter is significantly larger than for the former. The source size for the secondaries,  is similar but in general larger than the size for the whole pion sample. To further characterize the pion source, we also simulate the effects of a non-ideal detector introducing a momentum smearing parameter, representing the minimum pair momentum and thus a maximum source size that can be resolved. The values of the correlation function intercept parameter are therefore modified from the values they attain for the perfect detector case. Using  the core-halo picture of the source, we show that the values of the intercept parameter are influenced by the presence of a significant fraction of core pions coming from the decay of long-lived but slow-moving resonances. These findings serve as a benchmark to compare with future Monte Carlo studies that consider an Equation of State and thus allow for a phase transition within the studied energy domain.

\end{abstract}

\section{Introduction}

Two-particle correlation studies have become a prime tool to determine the size and lifetime, as well as to infer the global properties, of the strongly interacting systems produced in relativistic heavy-ion collisions ~\cite{Boal:1990yh,Weiner:1999th,Lednicky:2005af,Lisa:2008gf,Csorgo:2005gd,Lisa:2005dd,Kisiel:2011jt,Lokos:2022exf,Janik:2018ghw,Ayala:2021zst}. Since the two-particle correlation function is related to the Fourier transform of the spatio-temporal component of the phase-space density of the emitting source, measurements of the correlation function provide access in particular to the space-time features of this source. Pions are by far the most abundant particles produced in these collisions. Therefore, it is common to perform correlation studies by experimentally measuring the two-pion correlation function. The technique is closely related to the photon intensity interferometry measurements introduced by Hanbury-Brown and Twiss to determine the size of stellar objects~\cite{HanburyBrown:1956bqd,HanburyBrown:1954amm}. In the context of collisions of hadron systems, Goldhaber, Goldhaber, Lee and Pais~\cite{Goldhaber:1960sf} developed a similar technique to study the interaction region formed in these systems. Lednicky~\cite{Lednicky:2002fq} coined the term {\it femtoscopy} to emphasize  that the technique is in this case applied to measurements  at the fermi or femtometer scale. Femtoscopic studies have nowadays become a sophisticated and ever more precise tool, both experimentally as well as theoretically.

From the phenomenological point of view, the two-pion correlation function can be described in terms of a set of parameters that contain not only information about the space-time size of the pion emitting source but also about the kind of processes that drive pion production. These parameters include the average source size $R$ and the intercept $\lambda$ at vanishing relative momentum. Since particle production processes depend on the collision energy, it is expected that such parameters likewise evolve with energy. Of particular importance is to find whether the variation of these parameters show signals of criticality in the range of collision energies, where strongly interacting matter undergoes a transition from hadron to quark and gluon-dominated degrees of freedom. In this sense, femtoscopic studies provide also a tool for the exploration of the QCD phase diagram~\cite{Csanad:2020xbf,Bzdak:2019pkr}. Experimentally, a promising energy domain where signals of criticality can be found is planned to be scanned by the Multipurpose Detector (MPD)~\cite{MPD:2022qhn} at the Nuclotron-based Ion Collider fAcility (NICA), currently under construction at the Joint Institute for Nuclear Research (JINR).

When the correlation function is described as a function of the relative invariant pair momentum for a fixed average pair momentum,  it is common to assume that the shape of the correlation function can be parametrized in terms of a Gaussian. Since the Fourier transform of a Gaussian is also a Gaussian, this assumption provides a simple description of the space-time source from the information obtained in momentum space. However, in recent times, it has become clear that this simple parametrization is not adequate and that a better description is achieved if the correlation function is parametrized based on a source described by a symmetric Lévy distribution~\cite{Cimerman:2019hva,Nagy:2023zbg,Csorgo:2003uv}. This is based on the realization that particle sources may show a large tail in configuration space and thus a description in terms of a distribution containing only one characteristic length may not be appropriate. The symmetric L\'evy distribution is a generalization of the Gaussian distribution where the exponent $\alpha$ is called the L\'evy exponent. The case $\alpha=2$ corresponds to the Gaussian distribution and $\alpha = 1$ corresponds to a Cauchy distribution.

Two-particle correlation studies based on L\'evy shape fits have been performed to describe experimental data for a single, or simultaneously at most for a couple of collision energies, mainly in the large energy domain. These studies correspond to SPS~\cite{Porfy:2023yii}, RHIC~\cite{Kincses:2016jsr,Kincses:2017zlb,Lokos:2018qdl,Mukherjee:2023hrz,Bystersky:2005qx} and LHC~\cite{CMS:2023xyd} energies for nucleus-nucleus (A-A) collisions and also to LHC~\cite{Schegelsky:2018tit} energies for proton-proton collisions. A recent compilation in this energy range for the A-A case has been reported in Ref.~\cite{Csanad:2024hva}. However, these studies have not yet been performed, neither fitting experimental data nor at Monte Carlo (MC) level, for energies where the putative Critical End Point (CEP) is thought to exist. In this work we aim to set up these studies within the NICA energy range where the CEP could be found. We report on correlation studies using MC generated data for central Bi+Bi collisions, which is one of the beam species planed to be used at the startup of the MPD data taking. The collision energies considered are within the range that is planned to be explored by the MPD experiment: $\sqrt{s_{NN}} = 4.0$, 5.8, 7.7 and 9.2 GeV. For our present purposes, the MC generated data does not include an Equation of State (EoS) and thus it does not consider a possible phase transition at a given energy. In this sense, this work represents a benchmark study that we aim to use to compare with future studies that will consider an EoS in the MC generated data. To contrast the commonly used shapes that describe the correlation functions, these are fitted with Gaussian, exponential and symmetric L\'evy shapes and the fits are compared. We show that, as it happens for the large energy domain, in the considered range and for all the studied cases, the symmetric L\'evy shape provides the best fits. From the fits we extract the source parameters $R$ and $\lambda$, and in the case of the L\'evy fit, also $\alpha$ and find their dependence with the collision energy. We also separate the pion sample into pions coming from the decay of long-lived resonances, (secondary) and pions coming from direct processes (primary) and analyze the parameters describing their L\'evy shape fits. We show that, as expected, primary pions come from a source with a smaller average size than secondary pions. However, the source size for secondary pions, although similar, is in general larger than the source size for the whole pion sample. To further characterize the source, we then introduce in the MC simulation the effects of a smeared momentum determination, that can be translated into a minimum pair momentum resolution, to extract the fraction of pions that come from the core using the  {\it core-halo} picture of the source~\cite{Wiedemann:1996ig,Bolz:1992hc,Csorgo:1994in,Csorgo:1999sj}. This momentum smearing hampers resolving particles coming from the halo, and thus affects the extracted values of the intercept $\lambda$ in the correlation function. In this case, which simulates a real detector with a finite momentum resolution, the core-halo picture identifies the intercept $\lambda$ with the square of the fraction of pions coming from the core~\cite{Wiedemann:1996ig,PHENIX:2017ino}. We show that, for the energy range considered, the core still contains a significant fraction of pions that we interpret as coming from long-lived but slow-moving resonances and thus is not made up mainly from primary particles. We point out that future comparisons of MC studies that include an EoS with our results, can help to identify the emergence of particle production processes that change the relative abundance of primary and secondary pions in the core that in turn could signal critical behavior, which can be identified analyzing the properties of the L\'evy stability index $\alpha$ as the collision energy changes.

The work is organized as follows: In Sec.~\ref{II} we describe the generalities of the two-pion correlation function and the implementation of the MC simulation. In Sec.~\ref{III} we fit the MC generated correlation functions with Gaussian, exponential and symmetric L\'evy shapes and extract the correlation parameters. In Sec.~\ref{sec:core-halo}, we separate the source into two components; one corresponding to primary and another corresponding to secondary pions and analyze the two samples finding the average source sizes for each of these sets. We also show an example of the source image for a collision producing a pion source at a time where 90\% of the pions have attained freeze-out and compare the average size of the freeze-out hyper-surface with the L\'evy size, finding an excellent agreement. We then introduce a smeared momentum resolution to simulate the effects of a real detector. We find the values of the intercept parameter and use the core-halo picture to link this parameter with the square of the fraction of core pions. We find that the core contains a significant component of secondary pions that we interpret as coming from long-lived but slow-moving resonances. We finally summarize and conclude in Sec.~\ref{concl}.

\section{Monte Carlo simulations for the two-pion correlation function}\label{II}

\begin{table*}[t]
    \centering
    \begin{tabular}{c|c|c|c|c}
          & $R_{\text{inv}}$ [fm] & $\lambda$ & $\alpha$ & $\chi^2 /\, \text{ndf}$ \\ \hline \hline
         Exponential & $9.459 \pm 0.218$ & $1.18 \pm 0.019$ & -- & 14.849 \\ \hline
         Gaussian & $7.254 \pm 0.166$ & $0.919 \pm 0.018$ & -- & 38.301 \\ \hline
         L\'evy & $8.121 \pm 0.059$ & $1.05 \pm 0.006$ & $1.312 \pm 0.015$ & 0.835 \\ \hline
    \end{tabular}
    \caption{Parameters resulting from fits of the two-pion correlation function to exponential, Gaussian and L\'evy forms for the complete pion set obtained at $\sqrt{s_{NN}} = 5.8$ GeV.}
    \label{tab:Fit-all-58}
\end{table*}
The two-pion correlation function is defined as
\begin{equation}
    \label{eq:C2_Def}
    C_2 (p_1,p_2) = \frac{P_2 (p_1,p_2)}{P_1 (p_1)P_1(p_2)},
\end{equation}
where $P_1$ and $P_2$ are the single- and two-pion momentum distributions, respectively, and $p_1$ and $p_2$ are the four-momenta of each of the pions. Usually, the two-pion correlation function is analysed as a function of the relative four-momentum $q = p_1 - p_2$ and for fixed values of the average pair momentum $K = \tfrac{1}{2} (p_1 + p_2)$. The two-pion correlation function can also be related to the pion emitting source function in phase space, $S(x,p)$, by noting that if dynamical correlations, such as Coulomb or strong final state interactions are neglected, the single- and two-pion momentum distributions can be written as
\begin{IEEEeqnarray}{rCl}
    P_1(p) & = & \int d^4 x \: S(x,p) \left\vert \Psi_p (x)\right\vert^2, \\
    P_2(p_1,p_2) & = & \int d^4 x_1 d^4 x_2 \:S(x_1,p_1) S(x_2,p_2) \nonumber \\
    & & \times\: \left\vert \Psi_{p_1,p_2}(x_1,x_2)\right\vert^2 ,
\end{IEEEeqnarray}
where $\Psi_p$ and $\Psi_{p_1,p_2}$ are the single- and two-pion symmetrized wave functions. 

With the purpose of identifying possible non-Gaussian structures in the correlation function, which becomes difficult in a three-dimensional study, we perform a one-dimensional analysis in terms of the variable $q_{\text{inv}} = \sqrt{q_0^2-|\vec{q}|^2}$~\cite{PHENIX:2017ino}. Then, the two-pion correlation function can be written as
\begin{equation}
    C_2 (p_1,p_2) = 1+\Re\left[ \frac{\Tilde{S}(q_{\text{inv}},p_1) \Tilde{S}^*(q_{\text{inv}},p_2)}{\Tilde{S}(0,p_1) \Tilde{S}^*(0,p_2)} \right],
\end{equation}
where $\Tilde{S}$ is the Fourier transform of $S$. If the relative momentum between the particles is much smaller than the average pair momentum, then the two-pion correlation function can be written as~\cite{Wiedemann:1999qn,NA61SHINE:2023qzr}
\begin{equation}
    C_2(q_{\text{inv}},K) = 1+\frac{\left\vert \Tilde{S}(q_\text{inv},K)\right\vert^2}{\left\vert\Tilde{S}(0,K)\right\vert^2} ,
\end{equation}
which is a function of $q_{\mbox{\small{inv}}}$, for a fixed average pair momentum $K$. Since the simulations require a large amount of events to produce a statistically significant sample, hereby we consider all possible values of $K$ that contribute to a given $q_{\mbox{\small{inv}}}$. In this sense, our results have to be regarded
as describing an effective source that contains all possible sizes corresponding
to all values of the pair momenta. The resulting correlation function can thus be parametrized by different shapes, for instance Gaussian, exponential or L\'evy shapes~\cite{Csorgo:2003uv}.
From the previous equation, it can be seen that the two-pion correlation function can reach a maximum value of 2 at zero relative momentum, where the correlation function intercept, $C_2 (q_{\mbox{\small{inv}}}\to 0)$, is usually, denoted by $C_2 (q_{\mbox{\small{inv}}} \to 0) = 1 + \lambda$, with $\lambda$ also known as the chaoticity or intercept parameter. However, different effects such as final state interactions and a finite experimental resolution, can prevent the intercept parameter from reaching the value $1$ and can be understood in terms of the core-halo picture, whereby particles that come from the decays of long-lived resonances create a component of the source with a size that may not be resolved when the corresponding width of the pair momentum difference becomes smaller than the detector resolution. Accounting for this possibility the phase space emitting source can be modeled as consisting of two components $S = S_{\text{core}} + S_{\text{halo}}$, where each component has a Fourier transform and the core is composed of pions that come from primary processes. Notice that 
\begin{IEEEeqnarray}{rCl}
    N_{\text{core}} &=& \int\! d^4 x\: S_{\text{core}} (x) = \Tilde{S}_{\text{core}}(0), \\
    N_{\text{halo}}  &=& \int\! d^4 x\: S_{\text{halo}} (x) = \Tilde{S}_{\text{halo}}(0),
\end{IEEEeqnarray}
and hence, $\Tilde{S}(0) = N_{\text{core}} + N_{\text{halo}}$. Thus, for experimentally resolvable values of the relative momentum, it can be assumed that $\Tilde{S}(q_{\text{inv}}) \simeq \Tilde{S}_{\text{core}}(q_{\text{inv}})$. Therefore, the two-pion correlation function can be expressed as
\begin{equation}
    C_2 (q_{\text{inv}}) =1 + \left(\frac{N_{\text{core}}}{N_{\text{core}}+N_{\text{halo}}}\right)^2\!\frac{\left\vert\Tilde{S}_{\text{core}}(q_{\text{inv}})\right\vert^2}{\left\vert\Tilde{S}_{\text{core}}(0)\right\vert^2}.
\end{equation}
As a consequence, in the core-halo picture, we can identify for any given pair invariant  average momentum~\cite{Csorgo:2003uv,Wiedemann:1996ig}
\begin{equation}
    \label{eq:lambda}
    \lambda = \left(\frac{N_{\text{core}}}{N_{\text{core}}+N_{\text{halo}}}\right)^2.
\end{equation}

On the other hand, the previous definition of the two-pion correlation function, Eq.~\eqref{eq:C2_Def}, is barely used when dealing either with experimental or MC produced data. Instead of Eq.~\eqref{eq:C2_Def}, one usually defines the measured two-pion correlation function as~\cite{NA61SHINE:2023qzr,Zajc:1984vb}
\begin{equation}
    C_2 (q) = \frac{\mathcal{N}_B}{\mathcal{N}_A} \frac{A(q)}{B(q)},
\end{equation}
where $A(q)$ is the relative momentum distribution of pions created in the same event which contains the quantum statistical effects, and $B(q)$ is the relative momentum distribution of pions created in different events, which does not contain Bose-Einstein correlations. $\mathcal{N}_{A}$ and $\mathcal{N}_{B}$ are normalization factors for $A$ and $B$. 
\begin{figure}[b]
    \centering
    \includegraphics[width=0.475\textwidth]{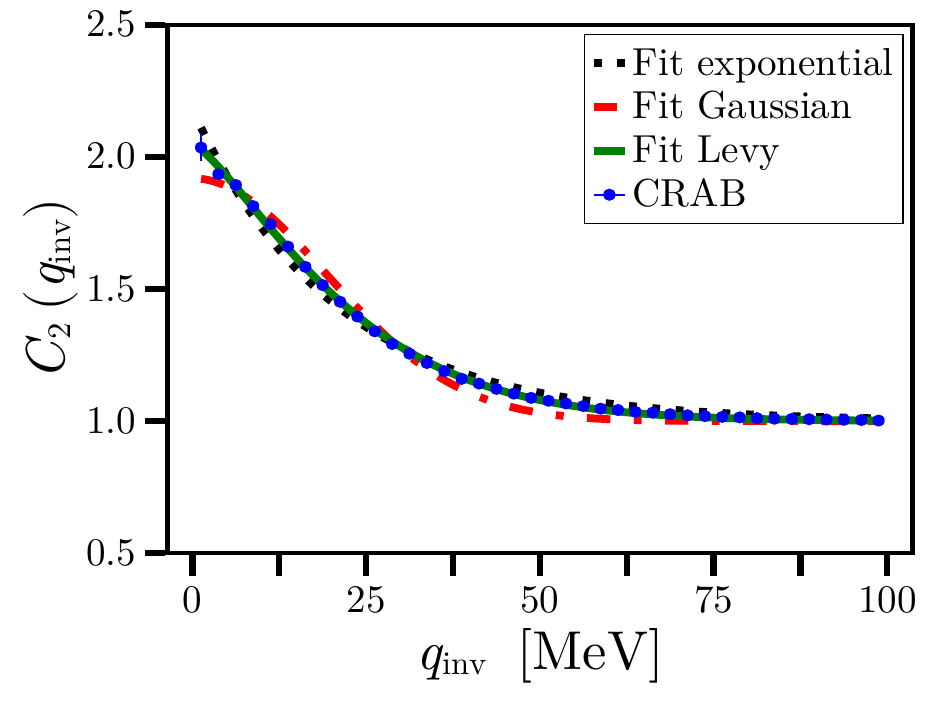}
    \caption{Two-pion correlation function for Bi+Bi collisions at $\sqrt{s_{NN}} = 5.8$ GeV, with impact parameter $b = 0-1$ fm. The dots represent the output of CRAB which is compared with exponential, Gaussian and L\'evy fits.}
    \label{fig:C2-all-58}
\end{figure}
\begin{figure}[b]
    \centering
    \subfloat[$t = 35$ fm]{{\includegraphics[width=0.475\textwidth]{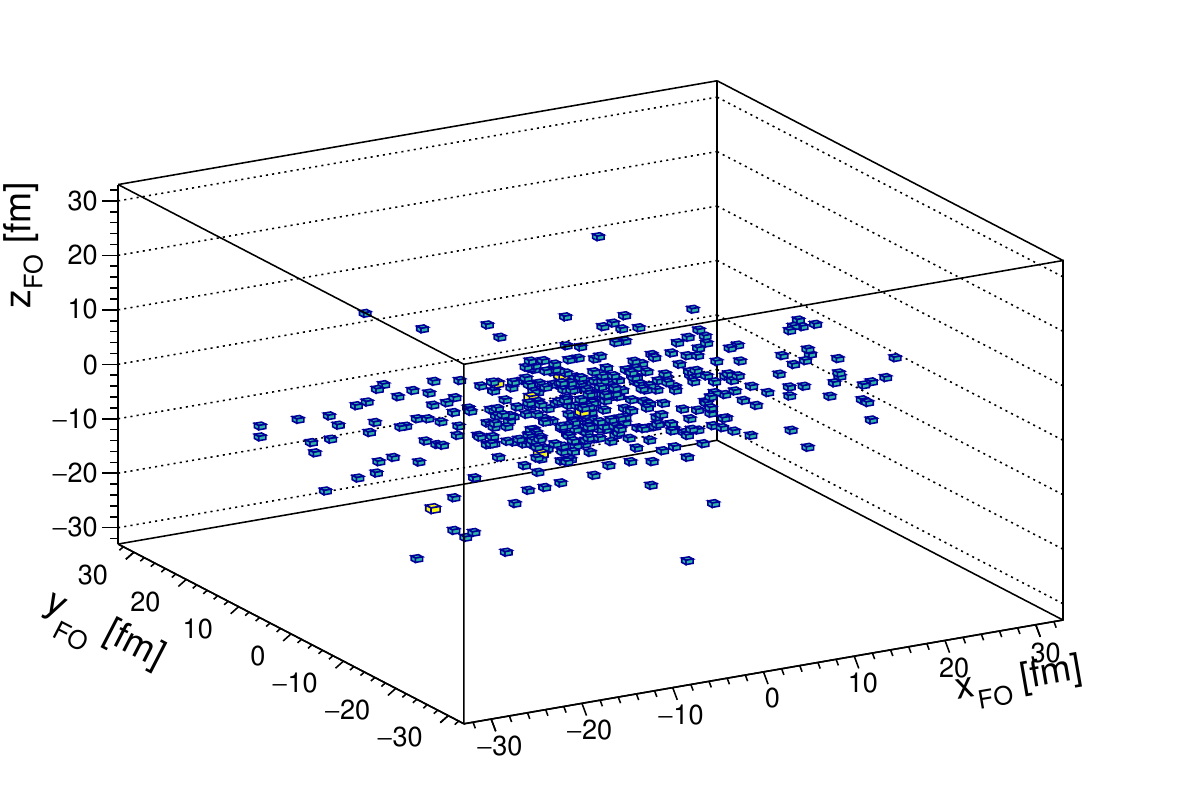}}}
    \qquad
    \subfloat[$t = 200$ fm]{{\includegraphics[width=0.475\textwidth]{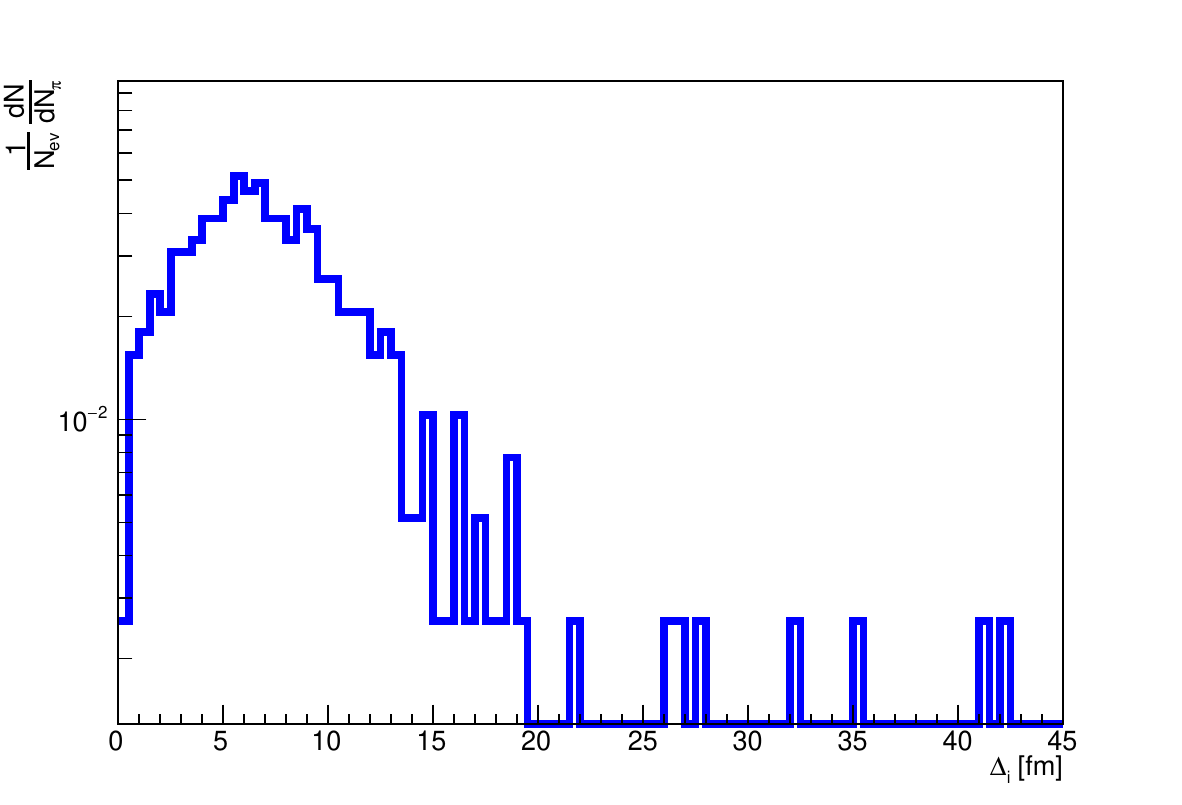}}}
    \caption{ (a) Example of the three-dimensional source image in the center-of-mass system for an event with $\sqrt{s_{NN}}=9.2$ GeV obtained from the spatial freeze-out coordinates given by UrQMD at time $t=35$ fm when 90\% of the charged pions have attained their freeze-out. The source is nearly spherical in coordinate space.  (b) One-dimensional distribution of $\Delta_i= \sqrt{(t_{\mbox{\tiny{FO}}}^2)_i - (|\vec{r}|_{\mbox{\tiny{FO}}}^2)_i}$ for each particle for a sample event obtained from the freeze-out coordinates given by UrQMD at time $t = 200$ fm. The average value is obtained as $\Delta = 9.19$ fm.}
    \label{fig:FOImage}
\end{figure}

In this work, we obtain the relative momentum distributions by means of MC simulations of relativistic heavy-ion collisions, from the Ultra-relativistic Quantum Molecular Dynamics Model (UrQMD)~\cite{Bass:1998ca,Bleicher:1999xi} in the cascade mode, which is used to simulate five million Bi+Bi central collisions at different collision energies within the NICA range ($\sqrt{s_{NN}} = 4.0$, 5.8, 7.7 and 9.2 GeV)  with the simulation stopped at a time 200 fm. Since UrQMD does not include quantum statistical correlations~\cite{Li:2012np,Ermakov:2017knq}, these are added by the formalism included in the ``correlation after-burner'' (CRAB) analyzing program~\cite{Pratt:1994uf}, which uses the phase space distributions at their freeze-out positions to implement correlation weights. CRAB performs a boost for
each of the pion pair momenta to the pair center of mass frame, where the pair
wave function is easier to symmetrize, to then produce the quantum correlation. This is expressed as a function
of the relative momentum which can then in turn be written in terms of $q_{\mbox{\small{inv}}}$.

\section{Correlation function fits and parameters}\label{III}

In general, it is assumed that the phase space source distribution, $S$, can be factorized into a space-time distribution and momentum distribution. The Fourier transform of the space-time part, which is often referred to as the characteristic function, is assumed to be an analytic function around zero relative momentum and its second order Taylor expansion characterizes its behavior, even for large values of $q_{\text{inv}}$~\cite{Csorgo:2003uv}. Thus, the two-pion correlation function can be approximately written as
\begin{table*}[ht]
    \centering
    \scalebox{0.85}{\begin{tabular}{c|c|c|c|c}
         & $R_{\text{inv}}$ [fm] & $\lambda$ & $\alpha$ & $\chi^2 / \, \text{ndf}$ \\ \hline \hline
         Primary & $3.516 \pm 0.014$ & $0.982 \pm 0.004$ & $1.863 \pm 0.021$ & 0.353  \\ \hline
         Secondary & $8.402 \pm 0.065$ & $1.066 \pm 0.007$ & $1.31 \pm 0.016$ & 0.321 \\ \hline
    \end{tabular}}
    \caption{Parameters resulting from fits of the two-pion correlation function to a L\'evy form for primary and secondary pions obtained at $\sqrt{s_{NN}} = 5.8$ GeV.}
    \label{tab:Fit-separation-58}
\end{table*}

\begin{figure}[b]
    \centering
    \includegraphics[width=0.475\textwidth]{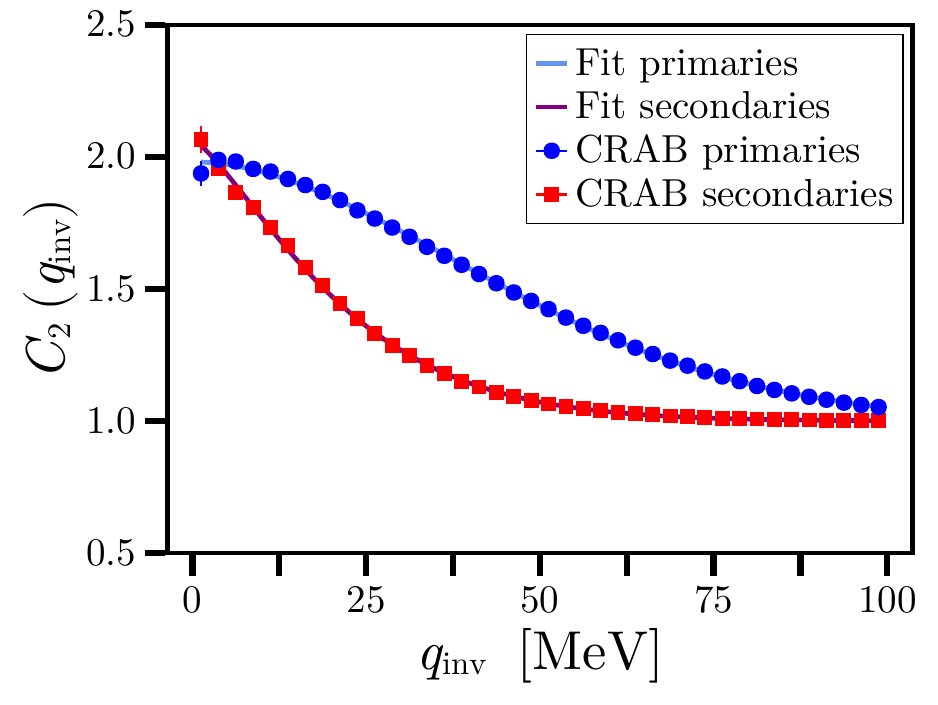}
    \caption{Two-pion correlation function of primary and secondary pions, produced in Bi+Bi collisions at $\sqrt{s_{NN}} = 5.8$ GeV, with impact parameter $b = 0-1$ fm. The dots represent the output of CRAB, while the solid lines represent the fit of L\'evy forms.}
    \label{fig:C2-separation-58}
\end{figure}
\begin{figure}[b]
    \centering
    \includegraphics[width=0.475\textwidth]{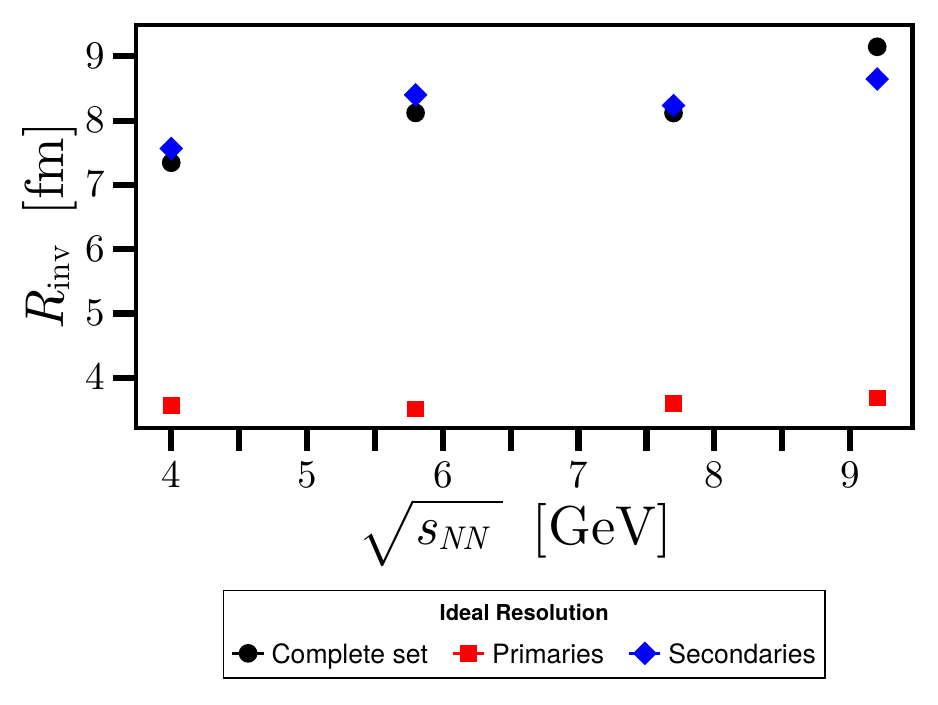}
    \caption{Source size $R_\text{inv}$ as a function of the collision energy, $\sqrt{s_{NN}}$, for the complete set of pions (black circles), primary pions (red rectangles) and secondary pions (blue diamonds), obtained from Bi+Bi collisions with impact parameter $b = 0-1$ fm. $R_\text{inv}$ is obtained from fits to a L\'evy form.}
    \label{fig:R-IR-dif-s}
\end{figure}
\begin{equation}
    \label{eq:C2Gauss}
    C_2 (q_{\text{inv}}) = 1 + \lambda \exp(-q_{\text{inv}} ^2 R_{\text{inv}}^2),
\end{equation}
where $R_{\text{inv}}$ is the characteristic size of the source. Eq.~\eqref{eq:C2Gauss} will be referred to as the Gaussian form. The previously mentioned assumptions can be translated into the stochastic nature of the several (independent) pion emission process. If one assumes that there are many independent processes that shift the emission position and that the final production point is a sum of many, similarly distributed, random shifts whose variance is finite, then according to the Central Limit Theorem, the distribution tends to a Gaussian. However, if the processes are characterized by large fluctuations originating power-like tails and a non-analytic behaviour of the characteristic function, then the limiting distribution is not a Gaussian but instead a L\'evy distribution. A special case of this distribution is the one named \textit{symmetric stable} L\'evy distribution. Utilizing such a distribution
for the source function, the correlation function is written as
\begin{equation}
    \label{eq:C2Levy}
    C_2 (q_{\text{inv}}) = 1 + \lambda \exp(-\vert q_{\text{inv}} R_{\text{inv}}\vert^\alpha),
\end{equation}\\
where $\alpha$ is called the index of stability and can be related to the correlation critical exponent of QCD~\cite{Csorgo:2008ayr,Csorgo:2004sr,Csorgo:2005it}. Equation~\eqref{eq:C2Levy} will be referred to as the L\'evy form of the correlation function.
As a last example of the different functions that can describe the two-pion correlation function, we consider the exponential shape
\begin{equation}
   \label{eq:C2Lorentz}
   C_2 (q_{\text{inv}}) = 1 + \lambda \exp(-\left|q_\text{inv} R_\text{inv}\right|),
\end{equation}
Equation~\eqref{eq:C2Lorentz} will be referred to as the exponential form of the correlation function.
\begin{figure}[b]
    \centering
    \includegraphics[width=0.475\textwidth]{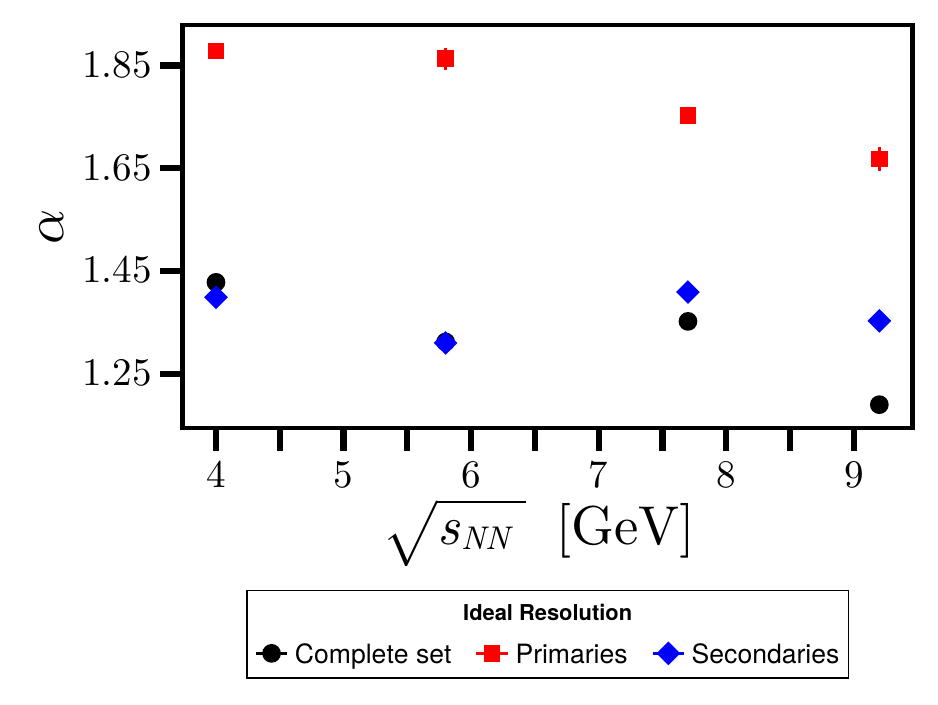}
    \caption{L\'evy index of stability $\alpha$ as a function of the collision energy, $\sqrt{s_{NN}}$, for the complete set of pions (black circles), primary pions (red rectangles) and secondary pions (blue diamonds), obtained from Bi+Bi collisions with impact parameter $b = 0-1$ fm. $\alpha$ is obtained from fits to a L\'evy form.}
    \label{fig:alpha-IR-dif-s}
\end{figure}

\section{\label{sec:core-halo}{Source features}
}
If the source can be thought of as consisting of two components, one made of pions coming from primary and another one from secondary processes, on average the distance from the center of the fireball, where these particles are produced, is different. This is the core-halo picture of particle production. In this section we test whether the simulated two-particle correlation functions can be analyzed separating the pion sample into primary and secondary pions, with the former coming from the core and the latter from the halo. As we proceed to show, this does not happen. To see this, we first study the case of a perfect resolution detector, and show that for the collision energies studied, although primary pions do indeed come from a small size source, the source size for secondary pions is similar albeit in general larger than the overall source size, which may seem puzzling. The picture is further elucidated when introducing a finite momentum resolution. As we also show, the fraction of pions coming from the core contains a significant portion of secondary pions which we attribute to the decay product of long-lived but slow-moving resonances. In this work we include all the resonances available in UrQMD. However the resonances that give rise to most of the secondary pions are $\rho$, $\Delta$, $\omega$, $K^*$, $N(1440)$, $\Delta(1600)$, $\Delta(1700)$, $\Delta(1950)$, $\Sigma$, $a_1$ and $\rho(1700)$~\cite{Bolz:1992hc}. Most of these resonances have not such a long life-time. However they are produced all over the life-time of the system. It is in this sense that we consider them to be \lq\lq long-lived". The picture that emerges is that the overall source size is the average between a small size core, containing a large population of secondaries, and a larger size halo. We extract the corresponding fireball parameters from fits to the central value of the correlation function in each bin. The rationale is to simulate the case of a very large statistical sample. We study the evolution of these parameters as a function of the collision energy for four cases: $\sqrt{s_{NN}}=4,\ 5.8,\ 7.7$ and $9.2$ GeV. In all cases the analyses are made for 5$\times 10^6$ central (impact parameter $b = 0-1$ fm) UrQMD Bi+Bi collisions to which CRAB is then applied.

\subsection{Ideal resolution case}
\begin{table*}[ht]
    \centering
    \begin{tabular}{c|c|c|c|c}
         & $R_{\text{inv}}$ [fm] & $\lambda$ & $\alpha$ & $\chi^2 / \, \text{ndf}$ \\ \hline \hline
          w.
          smearing  & $6.238 \pm 0.046$ & $0.632 \pm 0.004$ & $1.502 \pm 0.022$ & 0.912\\ \hline
         wo. smearing & $8.121 \pm 0.059$ & $1.05 \pm 0.006$ & $1.312 \pm 0.015$ & 0.835 \\ \hline
    \end{tabular}
    \caption{Results of the fit to the two-pion correlation function with a L\'evy form accounting for a finite resolution of the detector, setting a smearing of 10 MeV, for the complete pion set obtained at $\sqrt{s_{NN}} = 5.8$ GeV.}
    \label{tab:Fit-smear-58}
\end{table*}
\begin{figure}[b]
    \centering\includegraphics[width=0.475\textwidth]{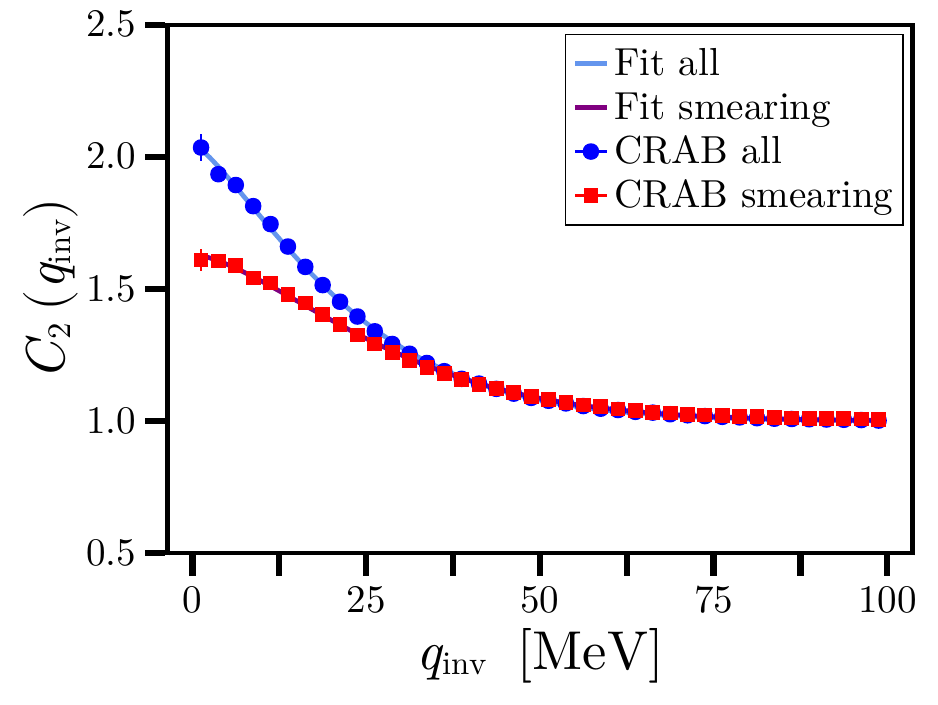}
    \caption{Two-pion correlation function of Bi+Bi collisions at $\sqrt{s_{NN}} = 5.8$ GeV, with impact parameter $b = 0-1$ fm. The blue dots represent the same output of CRAB of Figure~\ref{fig:C2-all-58}, while the red dots include the finite resolution effect of MPD, with a smearing of 10 MeV. Solid lines represent the fit to a L\'evy form.}
    \label{fig:C2-all-smearing-58}
\end{figure}

Figure~\ref{fig:C2-all-58} shows as an example the case for $\sqrt{s_{NN}} = 5.8$ GeV. The resulting parameters from fits to Gaussian, exponential and L\'evy shapes are shown in Table~\ref{tab:Fit-all-58}. Notice that the fit that better describes the  correlation function is obtained with the L\'evy shape. For all the studied shapes, from the fits we should obtain $\lambda = 1$. Nevertheless, the obtained values slightly deviate from 1 due to the limited statistics. The
deviation from 1 within errors is an indication of how good the sample is for statistical purposes. By increasing the statistics $\lambda$ should tend to 1 and the error bar become smaller.

The results of the source size obtained from the fit to the L\'evy form for the complete set of pions can be compared to the source image obtained from the freeze-out coordinates $(t_{\mbox{\tiny{FO}}},\vec{r}_{\mbox{\tiny{FO}}})$ given by UrQMD. An example of the three-dimensional source image is shown in Fig.~\ref{fig:FOImage}. The image is obtained in the center-of-mass system for a collision energy with $\sqrt{s_{NN}}=9.2$ GeV from the spatial freeze-out coordinates given by UrQMD at time $t=35$ fm when 90\% of the charged pions have frozen out. The source is nearly spherical in coordinate space. A long tail with a small density of frozen out pions can also be inferred from this image. The average space-time interval of the pion freeze-out coordinates $\Delta = \sqrt{\langle(t{\mbox{\tiny{FO}}}^2)_i - (|\vec{r}|_{\mbox{\tiny{FO}}}^2)_i\rangle}$ coincides, within a few percent, with the L\'evy fit value. We interpret this as the average width of the freeze-out hyper-surface, divided by a correction factor that accounts for the three-dimensional nature of this hyper-surface needed to compare with the one-dimensional L\'evy analysis.

To explore the origin of pions that populate the core and the halo, we first perform the separation of the sample into primary and secondary pions. We accomplish this using the UrQMD parent process identification, whereby secondary pions are those produced as a result of decays of long-lived resonances. Figure~\ref{fig:C2-separation-58} shows the separate contribution from primary and secondary pions to the the total correlation function shown in Fig.~\ref{fig:C2-all-58}. Since the best fit is obtained using a L\'evy form, hereafter we consider only the parameters obtained from this kind of fit,  
which are shown in Table~\ref{tab:Fit-separation-58}. The results are consistent with the picture whereby primary pions come mainly from the core, since their source has a size significantly smaller than that of secondary pions. 

\begin{figure}[b]
    \centering   \includegraphics[width=0.475\textwidth]{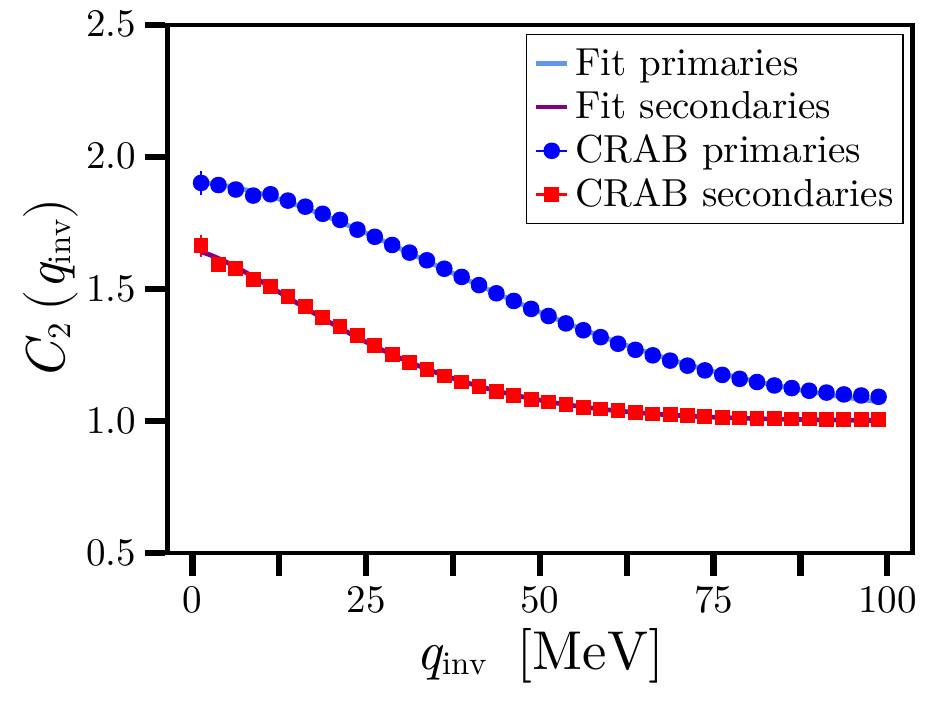}
    \caption{Two-pion correlation function of primary and secondary pions, produced in Bi+Bi collisions at $\sqrt{s_{NN}} = 5.8$ GeV, with impact parameter $b = 0-1$ fm. The dots represent the output of CRAB with a smearing of 10 MeV, while the solid lines represent the fit of L\'evy forms.}
    \label{fig:C2-separation-smearing-58}
\end{figure}


Figures~\ref{fig:R-IR-dif-s} and~\ref{fig:alpha-IR-dif-s} show the evolution of the source radii and of the L\'evy index of stability, respectively, with the collision energy for the complete, primary and secondary sets of pions. Notice that for most of the studied energies the following hierarchy of  source size parameters holds: $R_{\text{inv, prim}} < R_{\text{inv, all}} < R_{\text{inv, second}}$. except for the largest energy considered where we have instead $R_{\text{inv, prim}} <R_{\text{inv, second}}< R_{\text{inv, all}}$. This may seem somewhat confusing if one thinks that the larger portion of secondary pions should come from the halo, since they originate from long-lived resonances. We also notice that in general the following hierarchies for the intercept and L\'evy index: $\lambda_{\text{prim}} < \lambda_{\text{all}} < \lambda_{\text{second}}$ and $\alpha_{\text{second}} \lesssim \alpha_{\text{all}} < \alpha_{\text{prim}}$, and that the correlation function for primary pions has the closest behaviour to a Gaussian. For the whole sample of pions, there is a general tendency for $R_{\text{inv}}$ to increase as the collision energy increases. For the separate samples of primary and secondary pions, although $R_{\text{inv}}$ grows with the collision energy from the lowest to the largest energy considered, this growth is non-monotonic. The index $\alpha$ slightly decreases with the collision energy from the lowest to the largest energy considered but, except for the case of the primary pion sample, the overall decrease is non-monotonic. To summarize, for the perfect detector case, the general tendency for the source radii is to grow with the collision energy, while the L\'evy index of stability does not have a clear general tendency. The values of $R_{\text{inv}}$ for secondary pions are very similar but in general larger than those of the complete set of pions for the whole energy range; the opposite happens in general for the index $\alpha$.

\subsection{Finite resolution case}

\begin{table*}[t]
    \centering
    \begin{tabular}{c|c|c|c|c}
         & $R_{\text{inv}}$ [fm] & $\lambda$ & $\alpha$ & $\chi^2 / \, \text{ndf}$ \\ \hline \hline
         Primary & $3.426 \pm 0.012$ & $0.905 \pm 0.003$ & $1.709 \pm 0.016$ & 1.615 \\ \hline
         Secondary & $6.591 \pm 0.054$ & $0.647 \pm 0.005$ & $1.447 \pm 0.022$ & 0.259 \\ \hline
    \end{tabular}
    \caption{Results of the fit to the two-pion correlation function with a L\'evy form accounting for a finite resolution of the detector, setting a smearing of 10 MeV for primary and secondary pions at $\sqrt{s_{NN}} = 5.8$ GeV.}
    \label{tab:Fit-smear-separation-58}
\end{table*}

To account for finite resolution effects, recall that it has been reported that the NICA-MPD will have a minimum momentum resolution of about 1.5 \% for particles with total momentum around 0.2 GeV~\cite{Maevskiy:2020ank}. Therefore, the relative momentum resolution of MPD will be of about $\Delta_q=10$ MeV. This effect can be included in our studies by fixing the smearing parameter of CRAB to 10 MeV. This means that the momentum $p$ of a given particle will be assigned
the same value, provided it lies in the range $p\pm \Delta_q/2$, which in turn can be translated into a limit for the minimum pair momentum resolution. Our results show that for the studied energy range and for a non-perfect detector with a smearing, or equivalently, a track resolution $\Delta_q=10$ MeV, the resolvable size, in invariant variables, is of order 10 fm and the sample contains a non-resolvable halo which, as opposed to the case at higher collision energies, is basically a smooth continuation of the core, that is to say, it is not made of a very largely separated component. The existence of this non-resolvable component is signaled by a value $\lambda < 1$. By means of the Heisenberg uncertainty relation, this finite resolution therefore translates into the fact that source components whose characteristic size is larger than $R\gtrsim 1/\Delta_q$, cannot be resolved~\cite{Csorgo:1994in} and we hereby refer to these as the halo.

Figure~\ref{fig:C2-all-smearing-58} shows the effect of this finite resolution on the two-pion correlation function, shown in Figure~\ref{fig:C2-all-58}, together with a fit to a L\'evy form. The results of the fit are shown in Table~\ref{tab:Fit-smear-58}. Notice that the finite resolution has the effect of significantly diminish the source size (by about 23\%) and the value of the intercept (by about 39\%), while increasing the value of the L\'evy stability index (by about 14\%).

\begin{figure}[b]
    \centering
    \includegraphics[width=0.475\textwidth]{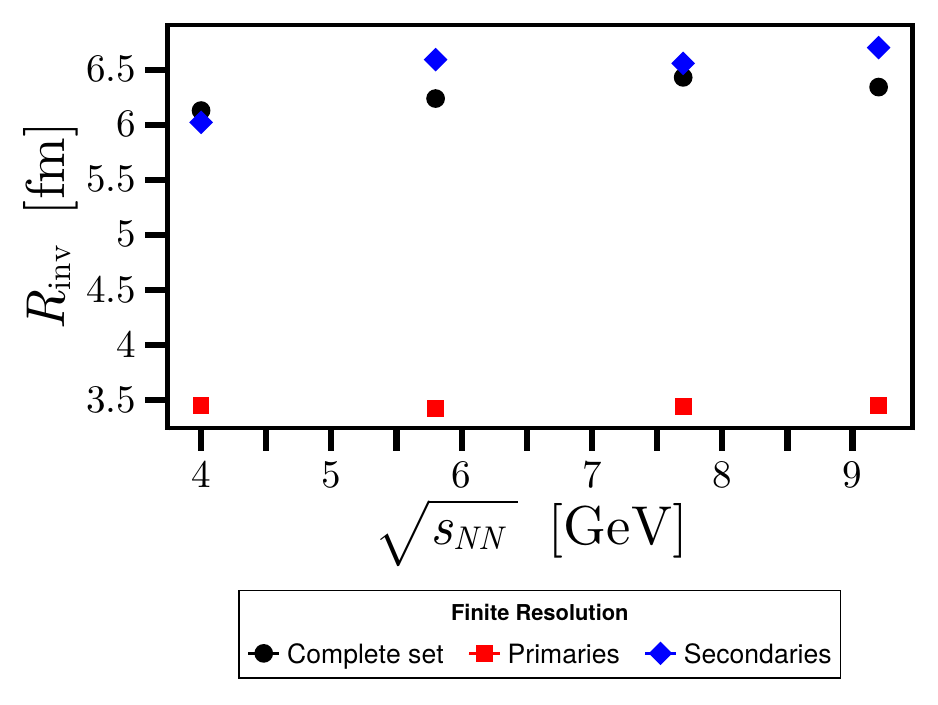}
    \caption{Source size $R_\text{inv}$ as a function of the collision energy, $\sqrt{s_{NN}}$, for the complete set of pions (black circles), primary pions (red rectangles) and secondary pions (blue diamonds), obtained from Bi+Bi collisions with impact parameter $b = 0-1$ fm and a smearing of 10 MeV. $R_\text{inv}$ is obtained from fits to a L\'evy form.}
    \label{fig:R-FR-dif-s}
\end{figure}

\begin{figure}[b!]
    \centering
    \includegraphics[width=0.475\textwidth]{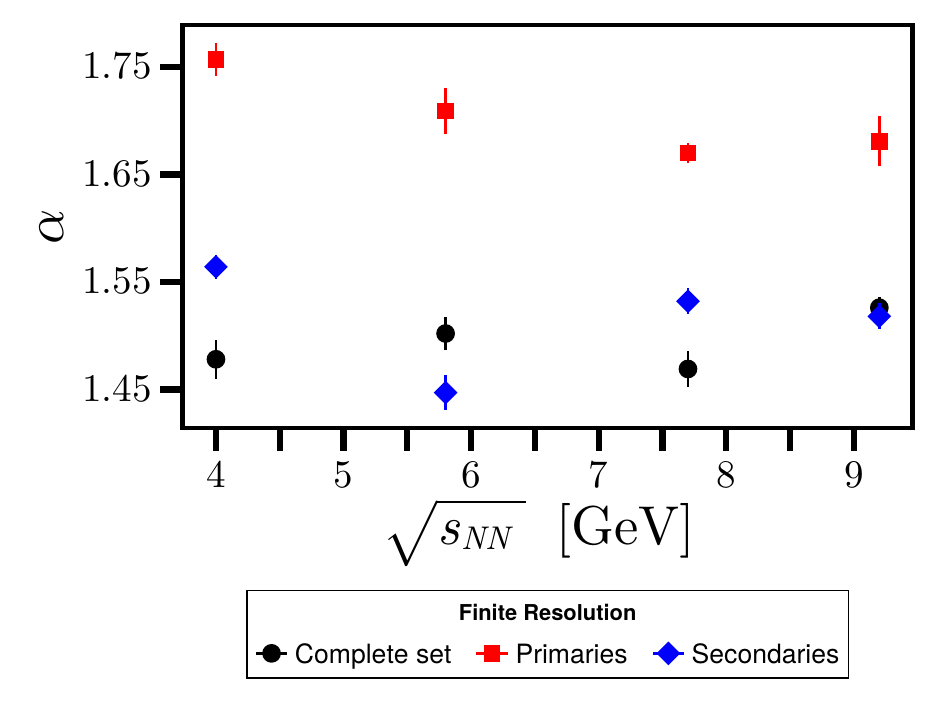}
    \caption{L\'evy index of stability $\alpha$ as a function of the collision energy, $\sqrt{s_{NN}}$, for the complete set of pions (black circles), primary pions (red rectangles) and secondary pions (blue diamonds), obtained from Bi+Bi collisions with impact parameter $b = 0-1$ fm and a smearing of 10 MeV. $\alpha$ is obtained from fits to a L\'evy form.}
    \label{fig:alpha-FR-dif-s}
\end{figure}
Figure~\ref{fig:C2-separation-smearing-58} shows the effect of a finite resolution (smearing of 10 MeV) on the separation of primary and secondary pions shown in Figure~\ref{fig:C2-separation-58}, together with a fit to a L\'evy form. The results of the fits are shown in Table~\ref{tab:Fit-smear-separation-58}. Notice that, once again, the finite resolution has the effect of diminishing the source size (by about 2\% for primary pions and about 21\%, for secondary pions) and the value of the intercept (by about 8\% for primary pions and about 39\% for secondary pions), while the value of the L\'evy index decreases for primary pions (by about 8\%) and increases for secondary pions (by about 10\%). This means that the effect of the finite resolution is of the same order for the set containing all the pions than for the set containing only the secondary pions, while the effect of a finite resolution barely affects the set of primary pions. 
\begin{table*}[t]
    \centering
    \begin{tabular}{c|c|c|c}
        $\sqrt{s_{NN}}$ [GeV] & $\lambda_{\text{all}}$ & $\lambda_{\text{prim}}$ & $\lambda_{\text{second}}$  \\ \hline \hline
        4.0 & $0.677 \pm 0.003$ & $0.907 \pm 0.002$ & $0.651 \pm 0.004$ \\ \hline
        5.8 & $0.632 \pm 0.004$ & $0.905 \pm 0.003$ & $0.647 \pm 0.005$ \\ \hline
        7.7 & $0.625 \pm 0.004$ & $0.9 \pm 0.003$ & $0.608 \pm 0.003$ \\ \hline
        9.2 & $0.595 \pm 0.007$ & $0.887 \pm 0.005$ & $0.602 \pm 0.003$ \\ \hline
    \end{tabular}
    \caption{Evolution of the intercept parameter $\lambda$ with the collision energy for the complete, primary and secondary sets of pions, obtained from fits to a L\'evy form accounting for a finite resolution of the detector with a smearing of 10 MeV.}
    \label{tab:lambda_sNN}
\end{table*}

\begin{table*}[t]
    \centering
    \begin{tabular}{c|c|c|c}
        $\sqrt{s_{NN}}$ [GeV] & All & Primaries & Secondaries  \\\hline\hline
        4.0 & 120 & 10 & 110 \\ \hline
        5.8 & 230 & 20 & 210 \\ \hline
        7.7 & 330 & 40 & 290 \\ \hline
        9.2 & 400 & 50 & 350 \\ \hline
    \end{tabular}
    \caption{Average charged pion multiplicity evolution with the collision energy for the complete, primary and secondary sets of pions.}
    \label{tab:Multiplicity}
\end{table*}

Figures~\ref{fig:R-FR-dif-s} and ~\ref{fig:alpha-FR-dif-s} show the evolution of the source radii and the L\'evy index, respectively, with the collision energy for the complete, primary and secondary sets of pions. The hierarchy of the fit parameters  for the invariant radii is $R_{\text{inv, prim, FR}} < R_{\text{inv, all, FR}} < R_{\text{inv, second, FR}}$, except for le lowest energy considered where we have instead $R_{\text{inv, prim, FR}} < R_{\text{inv, second, FR}} \lesssim R_{\text{inv, all, FR}}$. The hierarchy  for the intercept parameter becomes now $\lambda_{\text{all, FR}} < \lambda_{\text{second, FR}} < \lambda_{\text{prim, FR}}$, while it remains the same for the L\'evy index $\alpha_{\text{second, FR}} < \alpha_{\text{all, FR}} < \alpha_{\text{prim, FR}}$. Notice that for the set of primary pions, $R_{\text{inv}}$ is basically constant for the energy range considered, whereas $\alpha$ shows a moderate decrease with energy. This means that a finite momentum resolution has a small effect on the pions of primary origin. The values for $R_{\text{inv}}$ and $\alpha$ for the set of secondary pions are closer to the corresponding parameters when the whole pion sample is considered. This means that when no separation of the sample between primary and secondary pions is made, the full sample is dominated by the secondary pions. For the whole sample of pions, as well as for the set of secondary pions, $R_{\text{inv}}$ shows an overall tendency to increase with the collision energy, whereas for both of these samples, $\alpha$ is basically constant around the same value $\alpha\simeq 1.5$. 

Table~\ref{tab:lambda_sNN} shows the evolution of the intercept parameter $\lambda$ with respect to the collision energy for the complete set of pions, as well as for the primary and secondary pions. Notice that for all the cases, the value of $\lambda$ decreases as the energy increases. Nonetheless, for the complete set and the set of secondary pions, the decrease is around 10\%, whereas for the set of primary pions the decrease is only marginal and around 2\%. In the core-halo picture, the intercept parameter can be related to the square of the fraction of pions coming from the core. This implies, according to Eq.~\eqref{eq:lambda}, that between 77\% and 82\% of the pions should come from the core and only a smaller fraction should come from the halo.  According to our definition, pions of secondary origin are those coming from the decay or long-lived resonances. The results show that simulating a finite resolution detector with a smearing $\Delta q\sim$ 10 MeV, this fraction of pions of secondary origin together with the pions of primary origin come from a space-time region within $R_{\text{inv}}$, such that $\Delta q R_{\text{inv}}\lesssim 1$ implying that the space region that can be accessed is restricted to $R_{\text{inv}}\lesssim 20$ fm. We have verified from the simulations that indeed, about 20\% of the pions are produced in a region with a distance from the collision center larger than 20 fm. Therefore, the spatial region where about 20\% of secondary pions come from, corresponds to an average size $R_{\text{halo}}\gtrsim 20$ fm. This explains why, when considering a finite resolution detector, the core pions (as defined by the core-halo picture) do not contain this fraction of secondary pions  and thus, the average size for the source of secondaries decreases from $R_{\text{inv}}\sim 8$ fm to $R_\text{inv}\sim 6.5$ fm. Since the intercept parameter for secondaries decreases more than for the primaries as the energy increases, this means that for larger energies, long-lived resonances decay further away from the center, as expected, and thus contribute less to the population of core pions. From the UrQMD simulation, it is possible to directly identify that the average fraction of primary pions, which always populate the core, increases marginally with energy to be between 6\% and 12\%. Therefore, we can conclude that for lower energies, the core is mainly populated by pions from resonance decays and that this population slowly decreases as the collision energy increases. For completeness, and in order to fully characterize the pion sample, Table~\ref{tab:Multiplicity} shows the average multiplicity per event for the complete, the primary and the secondary sets of pions.

\section{Summary and Conclusions}\label{concl}

In the context of relativistic heavy-ion collisions within the NICA energy range, we have studied the evolution with collision energy of the parameters that describe the two-pion correlation function. We have performed MC simulations using the UrQMD event generator in the cascade mode, to produce 5$\times 10^6$ events for each considered energy. In each case, the quantum correlations are included using the CRAB analyzing code. No other source of correlations but the quantum ones have been included. We studied the correlation function as a function of the invariant relative pair momentum for a fixed value of the average pair momentum. To find the parameters that describe the correlation function we performed fits using Gaussian, exponential and symmetric L\'evy shapes. We have shown that, as is the case when considering larger energies, the L\'evy shape provides that best description of the correlations for the different settings and across the considered energy range. The most likely origin of
the L\'evy behavior is that the particle source has a large tail in configuration space
and thus a description in terms of a distribution containing only one characteristic
length is not appropriate. This is confirmed by our study of the three-dimensional source image. The pion sample is separated into its primary and secondary components. The latter is defined as the set of pions coming from the decay of long-lived resonances, mostly $\rho$, $\Delta$, $\omega$, $K^*$, $N(1440)$, $\Delta(1600)$, $\Delta(1700)$, $\Delta(1950)$, $\Sigma$, $a_1$ and $\rho(1700)$. We have shown that the source size for the sample of secondaries is similar but larger than that for the whole sample and significantly larger than the source size of the primaries. The intercept parameter exhibits the same hierarchy whereas the L\'evy index exhibits the opposite one. For the case of the primary pion sample, the L\'evy index shows an overall slight decrease with collision energy, however the secondary and full set of pions do not show a clear tendency with collision energy. We also notice that when restricting only to the set of primary pions, the correlation function already requires $\alpha < 2$. Adding to the sample of primary pions the pions coming from resonance decays produces in general a further decrease of $\alpha$~\cite{Kincses:2022eqq}.

In order to obtain a more accurate picture of the space-time characteristics of the pion producing sources, we have simulated the case of a non-ideal detector introducing a smearing parameter in the CRAB code to mimic a minimum resolution for the determination of the relative pair momentum. From the uncertainty relation between momentum and position, this translates into a maximum source size from where pion pairs can be identified. This smearing produces that the intercept of the correlation function becomes smaller than 1. Within the core-halo picture, the impossibility to determine source sizes larger than the inverse of the smearing momentum can be turned into an advantage since the size of the intercept can be directly identified with the square of the fraction of pions coming from the core. Our results indicate that the core pion sample has a large component that comes from the decay of long-lived but slow-moving resonances, as well as a small component of pions coming from primary processes. The former decreases whereas the latter increases with collision energy. In this sense, the analysis of the relative abundance of pions in the core coming from resonance decays and from primary processes, as the collision energy changes, becomes more important as a tool to study signals of criticality within the NICA energy range, when future comparisons are made with results from similar analysis but using an event generator that includes a phase transition within the same energy range. Indeed, when additional particle producing processes within the core introduce extra sources of correlations with lengths of order of the size of the system, such as when an EoS is considered in the MC generator, it is expected that these leave an imprint that can show up in particular as a non monotonic evolution of the L\'evy index with collision energy~\cite{Kincses:2022eqq}. In this sense, or results represent a benchmark to use for comparison with future studies where the event generator includes and EoS. We are performing this kind of analysis based on the findings of this work and the results will soon be reported elsewhere.

\section*{Acknowledgements}

The authors are in debt to M. Csan\'ad and to T. Cs\"org\"o for useful conversations and suggestions. The authors also thank JINR for providing the use of their firmware for computations that were carried out on the basis of the HybriLIT heterogeneous computing platform (LIT, JINR) \url{http://hlit.jinr.ru} ~\cite{adam2018ecosystem}. Support for this work was received in part by UNAM-PAPIIT grant number IG100322 and by Consejo Nacional de Humanidades, Ciencia y Tecnolog\'ia grant numbers CF-2023-G-433, A1-S-7655 and A1-S-16215. S. B. L. acknowledges the financial support of a fellowship granted by Consejo Nacional de Humanidades, Ciencia y Tecnolog\'ia as part of the Sistema Nacional de Posgrados. This research was also partially supported by the Munich Institute for Astro-, Particle and BioPhysics (MIAPbP), which is funded by the Deutsche Forschungsgemeinschaft (DFG, German Research Foundation) under Germany´s Excellence Strategy – EXC-2094 – 390783311.

\bibliographystyle{apsrev4-1}

\bibliography{biblio}

\begin{thebibliography}{48}%
\makeatletter
\providecommand \@ifxundefined [1]{%
 \@ifx{#1\undefined}
}%
\providecommand \@ifnum [1]{%
 \ifnum #1\expandafter \@firstoftwo
 \else \expandafter \@secondoftwo
 \fi
}%
\providecommand \@ifx [1]{%
 \ifx #1\expandafter \@firstoftwo
 \else \expandafter \@secondoftwo
 \fi
}%
\providecommand \natexlab [1]{#1}%
\providecommand \enquote  [1]{``#1''}%
\providecommand \bibnamefont  [1]{#1}%
\providecommand \bibfnamefont [1]{#1}%
\providecommand \citenamefont [1]{#1}%
\providecommand \href@noop [0]{\@secondoftwo}%
\providecommand \href [0]{\begingroup \@sanitize@url \@href}%
\providecommand \@href[1]{\@@startlink{#1}\@@href}%
\providecommand \@@href[1]{\endgroup#1\@@endlink}%
\providecommand \@sanitize@url [0]{\catcode `\\12\catcode `\$12\catcode
  `\&12\catcode `\#12\catcode `\^12\catcode `\_12\catcode `\%12\relax}%
\providecommand \@@startlink[1]{}%
\providecommand \@@endlink[0]{}%
\providecommand \url  [0]{\begingroup\@sanitize@url \@url }%
\providecommand \@url [1]{\endgroup\@href {#1}{\urlprefix }}%
\providecommand \urlprefix  [0]{URL }%
\providecommand \Eprint [0]{\href }%
\providecommand \doibase [0]{http://dx.doi.org/}%
\providecommand \selectlanguage [0]{\@gobble}%
\providecommand \bibinfo  [0]{\@secondoftwo}%
\providecommand \bibfield  [0]{\@secondoftwo}%
\providecommand \translation [1]{[#1]}%
\providecommand \BibitemOpen [0]{}%
\providecommand \bibitemStop [0]{}%
\providecommand \bibitemNoStop [0]{.\EOS\space}%
\providecommand \EOS [0]{\spacefactor3000\relax}%
\providecommand \BibitemShut  [1]{\csname bibitem#1\endcsname}%
\let\auto@bib@innerbib\@empty
\bibitem [{\citenamefont {Boal}\ \emph {et~al.}(1990)\citenamefont {Boal},
  \citenamefont {Gelbke},\ and\ \citenamefont {Jennings}}]{Boal:1990yh}%
  \BibitemOpen
  \bibfield  {author} {\bibinfo {author} {\bibfnamefont {D.~H.}\ \bibnamefont
  {Boal}}, \bibinfo {author} {\bibfnamefont {C.~K.}\ \bibnamefont {Gelbke}}, \
  and\ \bibinfo {author} {\bibfnamefont {B.~K.}\ \bibnamefont {Jennings}},\
  }\href {\doibase 10.1103/RevModPhys.62.553} {\bibfield  {journal} {\bibinfo
  {journal} {Rev. Mod. Phys.}\ }\textbf {\bibinfo {volume} {62}},\ \bibinfo
  {pages} {553} (\bibinfo {year} {1990})}\BibitemShut {NoStop}%
\bibitem [{\citenamefont {Weiner}(2000)}]{Weiner:1999th}%
  \BibitemOpen
  \bibfield  {author} {\bibinfo {author} {\bibfnamefont {R.~M.}\ \bibnamefont
  {Weiner}},\ }\href {\doibase 10.1016/S0370-1573(99)00114-3} {\bibfield
  {journal} {\bibinfo  {journal} {Phys. Rept.}\ }\textbf {\bibinfo {volume}
  {327}},\ \bibinfo {pages} {249} (\bibinfo {year} {2000})},\ \Eprint
  {http://arxiv.org/abs/hep-ph/9904389} {arXiv:hep-ph/9904389} \BibitemShut
  {NoStop}%
\bibitem [{\citenamefont {Lednicky}(2006)}]{Lednicky:2005af}%
  \BibitemOpen
  \bibfield  {author} {\bibinfo {author} {\bibfnamefont {R.}~\bibnamefont
  {Lednicky}},\ }\href {\doibase 10.1016/j.nuclphysa.2006.06.040} {\bibfield
  {journal} {\bibinfo  {journal} {Nucl. Phys. A}\ }\textbf {\bibinfo {volume}
  {774}},\ \bibinfo {pages} {189} (\bibinfo {year} {2006})},\ \Eprint
  {http://arxiv.org/abs/nucl-th/0510020} {arXiv:nucl-th/0510020} \BibitemShut
  {NoStop}%
\bibitem [{\citenamefont {Lisa}\ and\ \citenamefont
  {Pratt}(2010)}]{Lisa:2008gf}%
  \BibitemOpen
  \bibfield  {author} {\bibinfo {author} {\bibfnamefont {M.~A.}\ \bibnamefont
  {Lisa}}\ and\ \bibinfo {author} {\bibfnamefont {S.}~\bibnamefont {Pratt}},\
  }\enquote {\bibinfo {title} {{Femtoscopically Probing the Freeze-out
  Configuration in Heavy Ion Collisions}},}\ in\ \href {\doibase
  10.1007/978-3-642-01539-7_21} {\emph {\bibinfo {booktitle} {{Relativistic
  Heavy Ion Physics}}}},\ \bibinfo {editor} {edited by\ \bibinfo {editor}
  {\bibfnamefont {R.}~\bibnamefont {Stock}}}\ (\bibinfo {year} {2010})\ \Eprint
  {http://arxiv.org/abs/0811.1352} {arXiv:0811.1352 [nucl-ex]} \BibitemShut
  {NoStop}%
\bibitem [{\citenamefont {Csorgo}(2006)}]{Csorgo:2005gd}%
  \BibitemOpen
  \bibfield  {author} {\bibinfo {author} {\bibfnamefont {T.}~\bibnamefont
  {Csorgo}},\ }\href {\doibase 10.1088/1742-6596/50/1/031} {\bibfield
  {journal} {\bibinfo  {journal} {J. Phys. Conf. Ser.}\ }\textbf {\bibinfo
  {volume} {50}},\ \bibinfo {pages} {259} (\bibinfo {year} {2006})},\ \Eprint
  {http://arxiv.org/abs/nucl-th/0505019} {arXiv:nucl-th/0505019} \BibitemShut
  {NoStop}%
\bibitem [{\citenamefont {Lisa}\ \emph {et~al.}(2005)\citenamefont {Lisa},
  \citenamefont {Pratt}, \citenamefont {Soltz},\ and\ \citenamefont
  {Wiedemann}}]{Lisa:2005dd}%
  \BibitemOpen
  \bibfield  {author} {\bibinfo {author} {\bibfnamefont {M.~A.}\ \bibnamefont
  {Lisa}}, \bibinfo {author} {\bibfnamefont {S.}~\bibnamefont {Pratt}},
  \bibinfo {author} {\bibfnamefont {R.}~\bibnamefont {Soltz}}, \ and\ \bibinfo
  {author} {\bibfnamefont {U.}~\bibnamefont {Wiedemann}},\ }\href {\doibase
  10.1146/annurev.nucl.55.090704.151533} {\bibfield  {journal} {\bibinfo
  {journal} {Ann. Rev. Nucl. Part. Sci.}\ }\textbf {\bibinfo {volume} {55}},\
  \bibinfo {pages} {357} (\bibinfo {year} {2005})},\ \Eprint
  {http://arxiv.org/abs/nucl-ex/0505014} {arXiv:nucl-ex/0505014} \BibitemShut
  {NoStop}%
\bibitem [{\citenamefont {Kisiel}(2011)}]{Kisiel:2011jt}%
  \BibitemOpen
  \bibfield  {author} {\bibinfo {author} {\bibfnamefont {A.}~\bibnamefont
  {Kisiel}} (\bibinfo {collaboration} {ALICE}),\ }\href {\doibase
  10.22323/1.154.0003} {\bibfield  {journal} {\bibinfo  {journal} {PoS}\
  }\textbf {\bibinfo {volume} {WPCF2011}},\ \bibinfo {pages} {003} (\bibinfo
  {year} {2011})}\BibitemShut {NoStop}%
\bibitem [{\citenamefont {L\"ok\"os}(2022)}]{Lokos:2022exf}%
  \BibitemOpen
  \bibfield  {author} {\bibinfo {author} {\bibfnamefont {S.}~\bibnamefont
  {L\"ok\"os}},\ }\href {\doibase 10.5506/APhysPolBSupp.15.3-A30} {\bibfield
  {journal} {\bibinfo  {journal} {Acta Phys. Polon. Supp.}\ }\textbf {\bibinfo
  {volume} {15}},\ \bibinfo {pages} {30} (\bibinfo {year} {2022})},\ \Eprint
  {http://arxiv.org/abs/2206.13952} {arXiv:2206.13952 [hep-ex]} \BibitemShut
  {NoStop}%
\bibitem [{\citenamefont {Janik}(2018)}]{Janik:2018ghw}%
  \BibitemOpen
  \bibfield  {author} {\bibinfo {author} {\bibfnamefont {M.~A.}\ \bibnamefont
  {Janik}} (\bibinfo {collaboration} {ALICE}),\ }in\ \href@noop {} {\emph
  {\bibinfo {booktitle} {{13th Workshop on Particle Correlations and
  Femtoscopy}}}}\ (\bibinfo {year} {2018})\ \Eprint
  {http://arxiv.org/abs/1811.02828} {arXiv:1811.02828 [hep-ex]} \BibitemShut
  {NoStop}%
\bibitem [{\citenamefont {Ayala}\ \emph {et~al.}(2022)\citenamefont {Ayala},
  \citenamefont {Bernal-Langarica},\ and\ \citenamefont
  {Villavicencio}}]{Ayala:2021zst}%
  \BibitemOpen
  \bibfield  {author} {\bibinfo {author} {\bibfnamefont {A.}~\bibnamefont
  {Ayala}}, \bibinfo {author} {\bibfnamefont {S.}~\bibnamefont
  {Bernal-Langarica}}, \ and\ \bibinfo {author} {\bibfnamefont
  {C.}~\bibnamefont {Villavicencio}},\ }\href {\doibase
  10.1103/PhysRevD.105.056001} {\bibfield  {journal} {\bibinfo  {journal}
  {Phys. Rev. D}\ }\textbf {\bibinfo {volume} {105}},\ \bibinfo {pages}
  {056001} (\bibinfo {year} {2022})},\ \Eprint
  {http://arxiv.org/abs/2111.05951} {arXiv:2111.05951 [hep-ph]} \BibitemShut
  {NoStop}%
\bibitem [{\citenamefont {Hanbury~Brown}\ and\ \citenamefont
  {Twiss}(1956)}]{HanburyBrown:1956bqd}%
  \BibitemOpen
  \bibfield  {author} {\bibinfo {author} {\bibfnamefont {R.}~\bibnamefont
  {Hanbury~Brown}}\ and\ \bibinfo {author} {\bibfnamefont {R.~Q.}\ \bibnamefont
  {Twiss}},\ }\href {\doibase 10.1038/1781046a0} {\bibfield  {journal}
  {\bibinfo  {journal} {Nature}\ }\textbf {\bibinfo {volume} {178}},\ \bibinfo
  {pages} {1046} (\bibinfo {year} {1956})}\BibitemShut {NoStop}%
\bibitem [{\citenamefont {Hanbury~Brown}\ and\ \citenamefont
  {Twiss}(1954)}]{HanburyBrown:1954amm}%
  \BibitemOpen
  \bibfield  {author} {\bibinfo {author} {\bibfnamefont {R.}~\bibnamefont
  {Hanbury~Brown}}\ and\ \bibinfo {author} {\bibfnamefont {R.~Q.}\ \bibnamefont
  {Twiss}},\ }\href {\doibase 10.1080/14786440708520475} {\bibfield  {journal}
  {\bibinfo  {journal} {Phil. Mag. Ser. 7}\ }\textbf {\bibinfo {volume} {45}},\
  \bibinfo {pages} {663} (\bibinfo {year} {1954})}\BibitemShut {NoStop}%
\bibitem [{\citenamefont {Goldhaber}\ \emph {et~al.}(1960)\citenamefont
  {Goldhaber}, \citenamefont {Goldhaber}, \citenamefont {Lee},\ and\
  \citenamefont {Pais}}]{Goldhaber:1960sf}%
  \BibitemOpen
  \bibfield  {author} {\bibinfo {author} {\bibfnamefont {G.}~\bibnamefont
  {Goldhaber}}, \bibinfo {author} {\bibfnamefont {S.}~\bibnamefont
  {Goldhaber}}, \bibinfo {author} {\bibfnamefont {W.-Y.}\ \bibnamefont {Lee}},
  \ and\ \bibinfo {author} {\bibfnamefont {A.}~\bibnamefont {Pais}},\ }\href
  {\doibase 10.1103/PhysRev.120.300} {\bibfield  {journal} {\bibinfo  {journal}
  {Phys. Rev.}\ }\textbf {\bibinfo {volume} {120}},\ \bibinfo {pages} {300}
  (\bibinfo {year} {1960})}\BibitemShut {NoStop}%
\bibitem [{\citenamefont {Lednicky}(2002)}]{Lednicky:2002fq}%
  \BibitemOpen
  \bibfield  {author} {\bibinfo {author} {\bibfnamefont {R.}~\bibnamefont
  {Lednicky}},\ }in\ \href {\doibase 10.1142/9789812704962_0005} {\emph
  {\bibinfo {booktitle} {{32nd International Symposium on Multiparticle
  Dynamics}}}}\ (\bibinfo {year} {2002})\ pp.\ \bibinfo {pages} {21--26},\
  \Eprint {http://arxiv.org/abs/nucl-th/0212089} {arXiv:nucl-th/0212089}
  \BibitemShut {NoStop}%
\bibitem [{\citenamefont {Csanad}(2020)}]{Csanad:2020xbf}%
  \BibitemOpen
  \bibfield  {author} {\bibinfo {author} {\bibfnamefont {M.}~\bibnamefont
  {Csanad}} (\bibinfo {collaboration} {PHENIX}),\ }\href {\doibase
  10.1088/1742-6596/1602/1/012009} {\bibfield  {journal} {\bibinfo  {journal}
  {J. Phys. Conf. Ser.}\ }\textbf {\bibinfo {volume} {1602}},\ \bibinfo {pages}
  {012009} (\bibinfo {year} {2020})},\ \Eprint
  {http://arxiv.org/abs/2007.04751} {arXiv:2007.04751 [nucl-ex]} \BibitemShut
  {NoStop}%
\bibitem [{\citenamefont {Bzdak}\ \emph {et~al.}(2020)\citenamefont {Bzdak},
  \citenamefont {Esumi}, \citenamefont {Koch}, \citenamefont {Liao},
  \citenamefont {Stephanov},\ and\ \citenamefont {Xu}}]{Bzdak:2019pkr}%
  \BibitemOpen
  \bibfield  {author} {\bibinfo {author} {\bibfnamefont {A.}~\bibnamefont
  {Bzdak}}, \bibinfo {author} {\bibfnamefont {S.}~\bibnamefont {Esumi}},
  \bibinfo {author} {\bibfnamefont {V.}~\bibnamefont {Koch}}, \bibinfo {author}
  {\bibfnamefont {J.}~\bibnamefont {Liao}}, \bibinfo {author} {\bibfnamefont
  {M.}~\bibnamefont {Stephanov}}, \ and\ \bibinfo {author} {\bibfnamefont
  {N.}~\bibnamefont {Xu}},\ }\href {\doibase 10.1016/j.physrep.2020.01.005}
  {\bibfield  {journal} {\bibinfo  {journal} {Phys. Rept.}\ }\textbf {\bibinfo
  {volume} {853}},\ \bibinfo {pages} {1} (\bibinfo {year} {2020})},\ \Eprint
  {http://arxiv.org/abs/1906.00936} {arXiv:1906.00936 [nucl-th]} \BibitemShut
  {NoStop}%
\bibitem [{\citenamefont {Abgaryan}\ \emph {et~al.}(2022)\citenamefont
  {Abgaryan} \emph {et~al.}}]{MPD:2022qhn}%
  \BibitemOpen
  \bibfield  {author} {\bibinfo {author} {\bibfnamefont {V.}~\bibnamefont
  {Abgaryan}} \emph {et~al.} (\bibinfo {collaboration} {MPD}),\ }\href
  {\doibase 10.1140/epja/s10050-022-00750-6} {\bibfield  {journal} {\bibinfo
  {journal} {Eur. Phys. J. A}\ }\textbf {\bibinfo {volume} {58}},\ \bibinfo
  {pages} {140} (\bibinfo {year} {2022})},\ \Eprint
  {http://arxiv.org/abs/2202.08970} {arXiv:2202.08970 [physics.ins-det]}
  \BibitemShut {NoStop}%
\bibitem [{\citenamefont {Cimerma\v{n}}\ \emph {et~al.}(2020)\citenamefont
  {Cimerma\v{n}}, \citenamefont {Plumberg},\ and\ \citenamefont
  {Tom\'a\v{s}ik}}]{Cimerman:2019hva}%
  \BibitemOpen
  \bibfield  {author} {\bibinfo {author} {\bibfnamefont {J.}~\bibnamefont
  {Cimerma\v{n}}}, \bibinfo {author} {\bibfnamefont {C.}~\bibnamefont
  {Plumberg}}, \ and\ \bibinfo {author} {\bibfnamefont {B.}~\bibnamefont
  {Tom\'a\v{s}ik}},\ }\href@noop {} {\bibfield  {journal} {\bibinfo  {journal}
  {Phys. Part. Nucl.}\ }\textbf {\bibinfo {volume} {51}},\ \bibinfo {pages}
  {282} (\bibinfo {year} {2020})},\ \Eprint {http://arxiv.org/abs/1909.07998}
  {arXiv:1909.07998 [nucl-th]} \BibitemShut {NoStop}%
\bibitem [{\citenamefont {Nagy}\ \emph {et~al.}(2023)\citenamefont {Nagy},
  \citenamefont {Purzsa}, \citenamefont {Csan\'ad},\ and\ \citenamefont
  {Kincses}}]{Nagy:2023zbg}%
  \BibitemOpen
  \bibfield  {author} {\bibinfo {author} {\bibfnamefont {M.}~\bibnamefont
  {Nagy}}, \bibinfo {author} {\bibfnamefont {A.}~\bibnamefont {Purzsa}},
  \bibinfo {author} {\bibfnamefont {M.}~\bibnamefont {Csan\'ad}}, \ and\
  \bibinfo {author} {\bibfnamefont {D.}~\bibnamefont {Kincses}},\ }\href
  {\doibase 10.1140/epjc/s10052-023-12161-y} {\bibfield  {journal} {\bibinfo
  {journal} {Eur. Phys. J. C}\ }\textbf {\bibinfo {volume} {83}},\ \bibinfo
  {pages} {1015} (\bibinfo {year} {2023})},\ \Eprint
  {http://arxiv.org/abs/2308.10745} {arXiv:2308.10745 [nucl-th]} \BibitemShut
  {NoStop}%
\bibitem [{\citenamefont {Csorgo}\ \emph {et~al.}(2004)\citenamefont {Csorgo},
  \citenamefont {Hegyi},\ and\ \citenamefont {Zajc}}]{Csorgo:2003uv}%
  \BibitemOpen
  \bibfield  {author} {\bibinfo {author} {\bibfnamefont {T.}~\bibnamefont
  {Csorgo}}, \bibinfo {author} {\bibfnamefont {S.}~\bibnamefont {Hegyi}}, \
  and\ \bibinfo {author} {\bibfnamefont {W.~A.}\ \bibnamefont {Zajc}},\ }\href
  {\doibase 10.1140/epjc/s2004-01870-9} {\bibfield  {journal} {\bibinfo
  {journal} {Eur. Phys. J. C}\ }\textbf {\bibinfo {volume} {36}},\ \bibinfo
  {pages} {67} (\bibinfo {year} {2004})},\ \Eprint
  {http://arxiv.org/abs/nucl-th/0310042} {arXiv:nucl-th/0310042} \BibitemShut
  {NoStop}%
\bibitem [{\citenamefont {Porfy}(2023)}]{Porfy:2023yii}%
  \BibitemOpen
  \bibfield  {author} {\bibinfo {author} {\bibfnamefont {B.}~\bibnamefont
  {Porfy}} (\bibinfo {collaboration} {NA61/SHINE}),\ }\href {\doibase
  10.3390/universe9070298} {\bibfield  {journal} {\bibinfo  {journal}
  {Universe}\ }\textbf {\bibinfo {volume} {9}},\ \bibinfo {pages} {298}
  (\bibinfo {year} {2023})},\ \Eprint {http://arxiv.org/abs/2306.08696}
  {arXiv:2306.08696 [nucl-ex]} \BibitemShut {NoStop}%
\bibitem [{\citenamefont {Kincses}(2017)}]{Kincses:2016jsr}%
  \BibitemOpen
  \bibfield  {author} {\bibinfo {author} {\bibfnamefont {D.}~\bibnamefont
  {Kincses}} (\bibinfo {collaboration} {PHENIX}),\ }\href {\doibase
  10.5506/APhysPolBSupp.10.627} {\bibfield  {journal} {\bibinfo  {journal}
  {Acta Phys. Polon. Supp.}\ }\textbf {\bibinfo {volume} {10}},\ \bibinfo
  {pages} {627} (\bibinfo {year} {2017})},\ \Eprint
  {http://arxiv.org/abs/1610.05025} {arXiv:1610.05025 [nucl-ex]} \BibitemShut
  {NoStop}%
\bibitem [{\citenamefont {Kincses}(2018)}]{Kincses:2017zlb}%
  \BibitemOpen
  \bibfield  {author} {\bibinfo {author} {\bibfnamefont {D.}~\bibnamefont
  {Kincses}} (\bibinfo {collaboration} {PHENIX}),\ }\href {\doibase
  10.3390/universe4010011} {\bibfield  {journal} {\bibinfo  {journal}
  {Universe}\ }\textbf {\bibinfo {volume} {4}},\ \bibinfo {pages} {11}
  (\bibinfo {year} {2018})},\ \Eprint {http://arxiv.org/abs/1711.06891}
  {arXiv:1711.06891 [nucl-ex]} \BibitemShut {NoStop}%
\bibitem [{\citenamefont {L\"ok\"os}(2018)}]{Lokos:2018qdl}%
  \BibitemOpen
  \bibfield  {author} {\bibinfo {author} {\bibfnamefont {S.}~\bibnamefont
  {L\"ok\"os}} (\bibinfo {collaboration} {PHENIX}),\ }in\ \href@noop {} {\emph
  {\bibinfo {booktitle} {{13th Workshop on Particle Correlations and
  Femtoscopy}}}}\ (\bibinfo {year} {2018})\ \Eprint
  {http://arxiv.org/abs/1811.09788} {arXiv:1811.09788 [nucl-ex]} \BibitemShut
  {NoStop}%
\bibitem [{\citenamefont {Mukherjee}(2023)}]{Mukherjee:2023hrz}%
  \BibitemOpen
  \bibfield  {author} {\bibinfo {author} {\bibfnamefont {A.}~\bibnamefont
  {Mukherjee}},\ }\href {\doibase 10.3390/universe9070300} {\bibfield
  {journal} {\bibinfo  {journal} {Universe}\ }\textbf {\bibinfo {volume} {9}},\
  \bibinfo {pages} {300} (\bibinfo {year} {2023})},\ \Eprint
  {http://arxiv.org/abs/2306.13668} {arXiv:2306.13668 [nucl-ex]} \BibitemShut
  {NoStop}%
\bibitem [{\citenamefont {Bystersky}(2006)}]{Bystersky:2005qx}%
  \BibitemOpen
  \bibfield  {author} {\bibinfo {author} {\bibfnamefont {M.}~\bibnamefont
  {Bystersky}} (\bibinfo {collaboration} {STAR}),\ }\href {\doibase
  10.1063/1.2197466} {\bibfield  {journal} {\bibinfo  {journal} {AIP Conf.
  Proc.}\ }\textbf {\bibinfo {volume} {828}},\ \bibinfo {pages} {533} (\bibinfo
  {year} {2006})},\ \Eprint {http://arxiv.org/abs/nucl-ex/0511053}
  {arXiv:nucl-ex/0511053} \BibitemShut {NoStop}%
\bibitem [{\citenamefont {Tumasyan}\ \emph {et~al.}(2024)\citenamefont
  {Tumasyan} \emph {et~al.}}]{CMS:2023xyd}%
  \BibitemOpen
  \bibfield  {author} {\bibinfo {author} {\bibfnamefont {A.}~\bibnamefont
  {Tumasyan}} \emph {et~al.} (\bibinfo {collaboration} {CMS}),\ }\href
  {\doibase 10.1103/PhysRevC.109.024914} {\bibfield  {journal} {\bibinfo
  {journal} {Phys. Rev. C}\ }\textbf {\bibinfo {volume} {109}},\ \bibinfo
  {pages} {024914} (\bibinfo {year} {2024})},\ \Eprint
  {http://arxiv.org/abs/2306.11574} {arXiv:2306.11574 [nucl-ex]} \BibitemShut
  {NoStop}%
\bibitem [{\citenamefont {Schegelsky}(2019)}]{Schegelsky:2018tit}%
  \BibitemOpen
  \bibfield  {author} {\bibinfo {author} {\bibfnamefont {V.~A.}\ \bibnamefont
  {Schegelsky}},\ }\href {\doibase 10.1134/S1547477119050261} {\bibfield
  {journal} {\bibinfo  {journal} {Phys. Part. Nucl. Lett.}\ }\textbf {\bibinfo
  {volume} {16}},\ \bibinfo {pages} {503} (\bibinfo {year} {2019})},\ \Eprint
  {http://arxiv.org/abs/1804.07153} {arXiv:1804.07153 [hep-ph]} \BibitemShut
  {NoStop}%
\bibitem [{\citenamefont {Csan\'ad}\ and\ \citenamefont
  {Kincses}(2024)}]{Csanad:2024hva}%
  \BibitemOpen
  \bibfield  {author} {\bibinfo {author} {\bibfnamefont {M.}~\bibnamefont
  {Csan\'ad}}\ and\ \bibinfo {author} {\bibfnamefont {D.}~\bibnamefont
  {Kincses}},\ }in\ \href {\doibase 10.3390/universe10020054} {\emph {\bibinfo
  {booktitle} {{52nd International Symposium on Multiparticle Dynamics}}}}\
  (\bibinfo {year} {2024})\ \Eprint {http://arxiv.org/abs/2401.01249}
  {arXiv:2401.01249 [hep-ph]} \BibitemShut {NoStop}%
\bibitem [{\citenamefont {Wiedemann}\ and\ \citenamefont
  {Heinz}(1997)}]{Wiedemann:1996ig}%
  \BibitemOpen
  \bibfield  {author} {\bibinfo {author} {\bibfnamefont {U.~A.}\ \bibnamefont
  {Wiedemann}}\ and\ \bibinfo {author} {\bibfnamefont {U.~W.}\ \bibnamefont
  {Heinz}},\ }\href {\doibase 10.1103/PhysRevC.56.3265} {\bibfield  {journal}
  {\bibinfo  {journal} {Phys. Rev. C}\ }\textbf {\bibinfo {volume} {56}},\
  \bibinfo {pages} {3265} (\bibinfo {year} {1997})},\ \Eprint
  {http://arxiv.org/abs/nucl-th/9611031} {arXiv:nucl-th/9611031} \BibitemShut
  {NoStop}%
\bibitem [{\citenamefont {Bolz}\ \emph {et~al.}(1993)\citenamefont {Bolz},
  \citenamefont {Ornik}, \citenamefont {Plumer}, \citenamefont {Schlei},\ and\
  \citenamefont {Weiner}}]{Bolz:1992hc}%
  \BibitemOpen
  \bibfield  {author} {\bibinfo {author} {\bibfnamefont {J.}~\bibnamefont
  {Bolz}}, \bibinfo {author} {\bibfnamefont {U.}~\bibnamefont {Ornik}},
  \bibinfo {author} {\bibfnamefont {M.}~\bibnamefont {Plumer}}, \bibinfo
  {author} {\bibfnamefont {B.~R.}\ \bibnamefont {Schlei}}, \ and\ \bibinfo
  {author} {\bibfnamefont {R.~M.}\ \bibnamefont {Weiner}},\ }\href {\doibase
  10.1103/PhysRevD.47.3860} {\bibfield  {journal} {\bibinfo  {journal} {Phys.
  Rev. D}\ }\textbf {\bibinfo {volume} {47}},\ \bibinfo {pages} {3860}
  (\bibinfo {year} {1993})}\BibitemShut {NoStop}%
\bibitem [{\citenamefont {Csorgo}\ \emph {et~al.}(1996)\citenamefont {Csorgo},
  \citenamefont {Lorstad},\ and\ \citenamefont {Zimanyi}}]{Csorgo:1994in}%
  \BibitemOpen
  \bibfield  {author} {\bibinfo {author} {\bibfnamefont {T.}~\bibnamefont
  {Csorgo}}, \bibinfo {author} {\bibfnamefont {B.}~\bibnamefont {Lorstad}}, \
  and\ \bibinfo {author} {\bibfnamefont {J.}~\bibnamefont {Zimanyi}},\ }\href
  {\doibase 10.1007/s002880050195} {\bibfield  {journal} {\bibinfo  {journal}
  {Z. Phys. C}\ }\textbf {\bibinfo {volume} {71}},\ \bibinfo {pages} {491}
  (\bibinfo {year} {1996})},\ \Eprint {http://arxiv.org/abs/hep-ph/9411307}
  {arXiv:hep-ph/9411307} \BibitemShut {NoStop}%
\bibitem [{\citenamefont {Csorgo}(2002)}]{Csorgo:1999sj}%
  \BibitemOpen
  \bibfield  {author} {\bibinfo {author} {\bibfnamefont {T.}~\bibnamefont
  {Csorgo}},\ }\href {\doibase 10.1556/APH.15.2002.1-2.1} {\bibfield  {journal}
  {\bibinfo  {journal} {Acta Phys. Hung. A}\ }\textbf {\bibinfo {volume}
  {15}},\ \bibinfo {pages} {1} (\bibinfo {year} {2002})},\ \Eprint
  {http://arxiv.org/abs/hep-ph/0001233} {arXiv:hep-ph/0001233} \BibitemShut
  {NoStop}%
\bibitem [{\citenamefont {Adare}\ \emph {et~al.}(2018)\citenamefont {Adare}
  \emph {et~al.}}]{PHENIX:2017ino}%
  \BibitemOpen
  \bibfield  {author} {\bibinfo {author} {\bibfnamefont {A.}~\bibnamefont
  {Adare}} \emph {et~al.} (\bibinfo {collaboration} {PHENIX}),\ }\href
  {\doibase 10.1103/PhysRevC.97.064911} {\bibfield  {journal} {\bibinfo
  {journal} {Phys. Rev. C}\ }\textbf {\bibinfo {volume} {97}},\ \bibinfo
  {pages} {064911} (\bibinfo {year} {2018})},\ \Eprint
  {http://arxiv.org/abs/1709.05649} {arXiv:1709.05649 [nucl-ex]} \BibitemShut
  {NoStop}%
\bibitem [{\citenamefont {Wiedemann}\ and\ \citenamefont
  {Heinz}(1999)}]{Wiedemann:1999qn}%
  \BibitemOpen
  \bibfield  {author} {\bibinfo {author} {\bibfnamefont {U.~A.}\ \bibnamefont
  {Wiedemann}}\ and\ \bibinfo {author} {\bibfnamefont {U.~W.}\ \bibnamefont
  {Heinz}},\ }\href {\doibase 10.1016/S0370-1573(99)00032-0} {\bibfield
  {journal} {\bibinfo  {journal} {Phys. Rept.}\ }\textbf {\bibinfo {volume}
  {319}},\ \bibinfo {pages} {145} (\bibinfo {year} {1999})},\ \Eprint
  {http://arxiv.org/abs/nucl-th/9901094} {arXiv:nucl-th/9901094} \BibitemShut
  {NoStop}%
\bibitem [{\citenamefont {Adhikary}\ \emph {et~al.}(2023)\citenamefont
  {Adhikary} \emph {et~al.}}]{NA61SHINE:2023qzr}%
  \BibitemOpen
  \bibfield  {author} {\bibinfo {author} {\bibfnamefont {H.}~\bibnamefont
  {Adhikary}} \emph {et~al.} (\bibinfo {collaboration} {NA61/SHINE}),\
  }\href@noop {} {\  (\bibinfo {year} {2023})},\ \Eprint
  {http://arxiv.org/abs/2302.04593} {arXiv:2302.04593 [nucl-ex]} \BibitemShut
  {NoStop}%
\bibitem [{\citenamefont {Zajc}\ \emph {et~al.}(1984)\citenamefont {Zajc} \emph
  {et~al.}}]{Zajc:1984vb}%
  \BibitemOpen
  \bibfield  {author} {\bibinfo {author} {\bibfnamefont {W.~A.}\ \bibnamefont
  {Zajc}} \emph {et~al.},\ }\href {\doibase 10.1103/PhysRevC.29.2173}
  {\bibfield  {journal} {\bibinfo  {journal} {Phys. Rev. C}\ }\textbf {\bibinfo
  {volume} {29}},\ \bibinfo {pages} {2173} (\bibinfo {year}
  {1984})}\BibitemShut {NoStop}%
\bibitem [{\citenamefont {Bass}\ \emph {et~al.}(1998)\citenamefont {Bass} \emph
  {et~al.}}]{Bass:1998ca}%
  \BibitemOpen
  \bibfield  {author} {\bibinfo {author} {\bibfnamefont {S.~A.}\ \bibnamefont
  {Bass}} \emph {et~al.},\ }\href {\doibase 10.1016/S0146-6410(98)00058-1}
  {\bibfield  {journal} {\bibinfo  {journal} {Prog. Part. Nucl. Phys.}\
  }\textbf {\bibinfo {volume} {41}},\ \bibinfo {pages} {255} (\bibinfo {year}
  {1998})},\ \Eprint {http://arxiv.org/abs/nucl-th/9803035}
  {arXiv:nucl-th/9803035} \BibitemShut {NoStop}%
\bibitem [{\citenamefont {Bleicher}\ \emph {et~al.}(1999)\citenamefont
  {Bleicher} \emph {et~al.}}]{Bleicher:1999xi}%
  \BibitemOpen
  \bibfield  {author} {\bibinfo {author} {\bibfnamefont {M.}~\bibnamefont
  {Bleicher}} \emph {et~al.},\ }\href {\doibase 10.1088/0954-3899/25/9/308}
  {\bibfield  {journal} {\bibinfo  {journal} {J. Phys. G}\ }\textbf {\bibinfo
  {volume} {25}},\ \bibinfo {pages} {1859} (\bibinfo {year} {1999})},\ \Eprint
  {http://arxiv.org/abs/hep-ph/9909407} {arXiv:hep-ph/9909407} \BibitemShut
  {NoStop}%
\bibitem [{\citenamefont {Li}\ \emph {et~al.}(2013)\citenamefont {Li},
  \citenamefont {Graef},\ and\ \citenamefont {Bleicher}}]{Li:2012np}%
  \BibitemOpen
  \bibfield  {author} {\bibinfo {author} {\bibfnamefont {Q.}~\bibnamefont
  {Li}}, \bibinfo {author} {\bibfnamefont {G.}~\bibnamefont {Graef}}, \ and\
  \bibinfo {author} {\bibfnamefont {M.}~\bibnamefont {Bleicher}},\ }\href
  {\doibase 10.1088/1742-6596/420/1/012039} {\bibfield  {journal} {\bibinfo
  {journal} {J. Phys. Conf. Ser.}\ }\textbf {\bibinfo {volume} {420}},\
  \bibinfo {pages} {012039} (\bibinfo {year} {2013})},\ \Eprint
  {http://arxiv.org/abs/1209.0042} {arXiv:1209.0042 [hep-ph]} \BibitemShut
  {NoStop}%
\bibitem [{\citenamefont {Ermakov}\ and\ \citenamefont
  {Nigmatkulov}(2017)}]{Ermakov:2017knq}%
  \BibitemOpen
  \bibfield  {author} {\bibinfo {author} {\bibfnamefont {N.}~\bibnamefont
  {Ermakov}}\ and\ \bibinfo {author} {\bibfnamefont {G.}~\bibnamefont
  {Nigmatkulov}},\ }\href {\doibase 10.1088/1742-6596/798/1/012055} {\bibfield
  {journal} {\bibinfo  {journal} {J. Phys. Conf. Ser.}\ }\textbf {\bibinfo
  {volume} {798}},\ \bibinfo {pages} {012055} (\bibinfo {year} {2017})},\
  \Eprint {http://arxiv.org/abs/1806.03550} {arXiv:1806.03550 [nucl-ex]}
  \BibitemShut {NoStop}%
\bibitem [{\citenamefont {Pratt}\ \emph {et~al.}(1994)\citenamefont {Pratt}
  \emph {et~al.}}]{Pratt:1994uf}%
  \BibitemOpen
  \bibfield  {author} {\bibinfo {author} {\bibfnamefont {S.}~\bibnamefont
  {Pratt}} \emph {et~al.},\ }\href {\doibase 10.1016/0375-9474(94)90614-9}
  {\bibfield  {journal} {\bibinfo  {journal} {Nucl. Phys. A}\ }\textbf
  {\bibinfo {volume} {566}},\ \bibinfo {pages} {103C} (\bibinfo {year}
  {1994})}\BibitemShut {NoStop}%
\bibitem [{\citenamefont {Csorgo}(2008)}]{Csorgo:2008ayr}%
  \BibitemOpen
  \bibfield  {author} {\bibinfo {author} {\bibfnamefont {T.}~\bibnamefont
  {Csorgo}},\ }\href {\doibase 10.22323/1.076.0027} {\bibfield  {journal}
  {\bibinfo  {journal} {PoS}\ }\textbf {\bibinfo {volume} {HIGH-PTLHC08}},\
  \bibinfo {pages} {027} (\bibinfo {year} {2008})},\ \Eprint
  {http://arxiv.org/abs/0903.0669} {arXiv:0903.0669 [nucl-th]} \BibitemShut
  {NoStop}%
\bibitem [{\citenamefont {Csorgo}\ \emph {et~al.}(2005)\citenamefont {Csorgo},
  \citenamefont {Hegyi}, \citenamefont {Novak},\ and\ \citenamefont
  {Zajc}}]{Csorgo:2004sr}%
  \BibitemOpen
  \bibfield  {author} {\bibinfo {author} {\bibfnamefont {T.}~\bibnamefont
  {Csorgo}}, \bibinfo {author} {\bibfnamefont {S.}~\bibnamefont {Hegyi}},
  \bibinfo {author} {\bibfnamefont {T.}~\bibnamefont {Novak}}, \ and\ \bibinfo
  {author} {\bibfnamefont {W.~A.}\ \bibnamefont {Zajc}},\ }\href@noop {}
  {\bibfield  {journal} {\bibinfo  {journal} {Acta Phys. Polon. B}\ }\textbf
  {\bibinfo {volume} {36}},\ \bibinfo {pages} {329} (\bibinfo {year} {2005})},\
  \Eprint {http://arxiv.org/abs/hep-ph/0412243} {arXiv:hep-ph/0412243}
  \BibitemShut {NoStop}%
\bibitem [{\citenamefont {Csorgo}\ \emph {et~al.}(2006)\citenamefont {Csorgo},
  \citenamefont {Hegyi}, \citenamefont {Novak},\ and\ \citenamefont
  {Zajc}}]{Csorgo:2005it}%
  \BibitemOpen
  \bibfield  {author} {\bibinfo {author} {\bibfnamefont {T.}~\bibnamefont
  {Csorgo}}, \bibinfo {author} {\bibfnamefont {S.}~\bibnamefont {Hegyi}},
  \bibinfo {author} {\bibfnamefont {T.}~\bibnamefont {Novak}}, \ and\ \bibinfo
  {author} {\bibfnamefont {W.~A.}\ \bibnamefont {Zajc}},\ }\href {\doibase
  10.1063/1.2197465} {\bibfield  {journal} {\bibinfo  {journal} {AIP Conf.
  Proc.}\ }\textbf {\bibinfo {volume} {828}},\ \bibinfo {pages} {525} (\bibinfo
  {year} {2006})},\ \Eprint {http://arxiv.org/abs/nucl-th/0512060}
  {arXiv:nucl-th/0512060} \BibitemShut {NoStop}%
\bibitem [{\citenamefont {Maevskiy}\ \emph {et~al.}(2021)\citenamefont
  {Maevskiy}, \citenamefont {Ratnikov}, \citenamefont {Zinchenko},\ and\
  \citenamefont {Riabov}}]{Maevskiy:2020ank}%
  \BibitemOpen
  \bibfield  {author} {\bibinfo {author} {\bibfnamefont {A.}~\bibnamefont
  {Maevskiy}}, \bibinfo {author} {\bibfnamefont {F.}~\bibnamefont {Ratnikov}},
  \bibinfo {author} {\bibfnamefont {A.}~\bibnamefont {Zinchenko}}, \ and\
  \bibinfo {author} {\bibfnamefont {V.}~\bibnamefont {Riabov}},\ }\href
  {\doibase 10.1140/epjc/s10052-021-09366-4} {\bibfield  {journal} {\bibinfo
  {journal} {Eur. Phys. J. C}\ }\textbf {\bibinfo {volume} {81}},\ \bibinfo
  {pages} {599} (\bibinfo {year} {2021})},\ \Eprint
  {http://arxiv.org/abs/2012.04595} {arXiv:2012.04595 [physics.ins-det]}
  \BibitemShut {NoStop}%
\bibitem [{\citenamefont {Kincses}\ \emph {et~al.}(2022)\citenamefont
  {Kincses}, \citenamefont {Stefaniak},\ and\ \citenamefont
  {Csan\'ad}}]{Kincses:2022eqq}%
  \BibitemOpen
  \bibfield  {author} {\bibinfo {author} {\bibfnamefont {D.}~\bibnamefont
  {Kincses}}, \bibinfo {author} {\bibfnamefont {M.}~\bibnamefont {Stefaniak}},
  \ and\ \bibinfo {author} {\bibfnamefont {M.}~\bibnamefont {Csan\'ad}},\
  }\href {\doibase 10.3390/e24030308} {\bibfield  {journal} {\bibinfo
  {journal} {Entropy}\ }\textbf {\bibinfo {volume} {24}},\ \bibinfo {pages}
  {308} (\bibinfo {year} {2022})},\ \Eprint {http://arxiv.org/abs/2201.07962}
  {arXiv:2201.07962 [hep-ph]} \BibitemShut {NoStop}%
\bibitem [{\citenamefont {Adam}\ \emph {et~al.}(2018)\citenamefont {Adam},
  \citenamefont {Bashashin}, \citenamefont {Belyakov}, \citenamefont
  {Kirakosyan}, \citenamefont {Matveev}, \citenamefont {Podgainy},
  \citenamefont {Sapozhnikova}, \citenamefont {Streltsova}, \citenamefont
  {Torosyan}, \citenamefont {Vala} \emph {et~al.}}]{adam2018ecosystem}%
  \BibitemOpen
  \bibfield  {author} {\bibinfo {author} {\bibfnamefont {G.}~\bibnamefont
  {Adam}}, \bibinfo {author} {\bibfnamefont {M.}~\bibnamefont {Bashashin}},
  \bibinfo {author} {\bibfnamefont {D.}~\bibnamefont {Belyakov}}, \bibinfo
  {author} {\bibfnamefont {M.}~\bibnamefont {Kirakosyan}}, \bibinfo {author}
  {\bibfnamefont {M.}~\bibnamefont {Matveev}}, \bibinfo {author} {\bibfnamefont
  {D.}~\bibnamefont {Podgainy}}, \bibinfo {author} {\bibfnamefont
  {T.}~\bibnamefont {Sapozhnikova}}, \bibinfo {author} {\bibfnamefont
  {O.}~\bibnamefont {Streltsova}}, \bibinfo {author} {\bibfnamefont
  {S.}~\bibnamefont {Torosyan}}, \bibinfo {author} {\bibfnamefont
  {M.}~\bibnamefont {Vala}},  \emph {et~al.},\ }in\ \href
  {https://ceur-ws.org/Vol-2267/638-644-paper-122.pdf} {\emph {\bibinfo
  {booktitle} {CEUR workshop proceedings}}},\ Vol.\ \bibinfo {volume} {2267}\
  (\bibinfo {year} {2018})\ pp.\ \bibinfo {pages} {638--644}\BibitemShut
  {NoStop}%
\end{thebibliography}%

\end{document}